\documentclass[paper,prd,reprint,nofootinbib,notitlepage,showpacs,showkeys]{revtex4-1}
\usepackage{graphicx}
\usepackage{amsmath}
\usepackage{amssymb}
\usepackage{dcolumn}
\usepackage{bm}
\usepackage{color}
\usepackage{dsfont}
\usepackage{epstopdf}

\newcommand \beq{\begin{eqnarray}}
\newcommand \eeq{\end{eqnarray}}

\begin{document}
\allowdisplaybreaks

\title{Bose-Einstein condensation and Silver Blaze property\\ from the two-loop $\Phi$-derivable approximation}

\author{Gergely Mark{\'o}}
\email{smarkovics@hotmail.com}
\affiliation{Centre de Physique Th{\'e}orique, Ecole Polytechnique, CNRS, 91128 Palaiseau Cedex, France.}

\author{Urko Reinosa}
\email{reinosa@cpht.polytechnique.fr}
\affiliation{Centre de Physique Th{\'e}orique, Ecole Polytechnique, CNRS, 91128 Palaiseau Cedex, France.}

\author{Zsolt Sz{\'e}p}
\email{szepzs@achilles.elte.hu}
\affiliation{MTA-ELTE Statistical and Biological Physics Research Group, H-1117 Budapest, Hungary.}


\begin{abstract}
We extend our previous investigation of the two-loop $\Phi$-derivable approximation to finite chemical potential $\mu$ and discuss Bose-Einstein condensation (BEC) in the case of a charged scalar field with $O(2)$ symmetry. We show that the approximation is renormalizable by means of counterterms which are independent of both the temperature and the chemical potential. We point out the presence of an additional skew contribution to the propagator as compared to the $\smash{\mu=0}$ case, which comes with its own gap equation (except at Hartree level). We solve this equation together with the field equation, and the usual longitudinal and transversal gap equations to find that the transition is second order, in agreement with recent lattice results to which we compare. We also discuss a general criterion an approximation should obey for the so-called Silver Blaze property to hold, and we show that any $\Phi$-derivable approximation at finite temperature and density obeys this criterion if one chooses a UV regularization that does not cut off the Matsubara sums.
\end{abstract}


\maketitle 

\section{Introduction}
A continuous effort is being invested in understanding non-perturbative phenomena at finite density using functional methods. The main motivation is the exploration of the phase diagram of strongly interacting matter on the baryon density vs temperature plane. Existing first principle or effective model calculations involve either lattice field theory simulations (for a review on Monte-Carlo results see \cite{FodorKatz:2009arxiv}, for more recent developments using the complex Langevin equation see \cite{Sexty:2013ica,Aarts:2013uxa,Sexty:2014inpress,Aarts:2014bwa} or more analytical functional methods, e.g. the Dyson-Schwinger approach \cite{Fischer:2012vc,Fischer:2014vxa}, the functional renormalization group \cite{Herbst:2013ail,Fischer:2013eca}, hard thermal loop calculations \cite{Haque:2014rua} or the n-particle irreducible (nPI) formalism \cite{Andersen:2006ys}. The interested reader can find a more exhaustive list of references in the review \cite{Fukushima:2010bq}.

The charged scalar field model, or $O(2)$ model, at finite density already exhibits some of the features and problems common to various theories including a chemical potential and for this reason, it is usually considered as a testing ground for method development. Depending on the values of the parameters, there could be a spontaneous symmetry breaking or Bose-Einstein condensation type phase transition and in this latter case the model exhibits the Silver Blaze property described in \cite{Cohen:2003kd}. On the lattice, the model also suffers from the sign problem, and it has therefore been used to test lattice methods which circumvent it \cite{Aarts:2008wh,Gattringer:2012df}. The charged scalar model has been discussed as the simplest model displaying Bose-Einstein condensation first in \cite{Kapusta:1981aa,Haber:1981fg,Haber:1981ts}, where both the free theory and the interacting theory including perturbative one-loop corrections were discussed. In \cite{Bernstein:1990kf} the Goldstone spectrum is analyzed using the one-loop partition function. The phase diagram has been discussed using the one-loop effective potential in \cite{Benson:1991nj}. In the same framework finite size corrections have been determined in \cite{Shiokawa:1998dk}, while the effects of the multiplicative anomaly have been studied in \cite{Elizalde:1997sw}. In \cite{Salasnich:2002pp} canonical thermal field theory was used to study the phase transition, while in \cite{Andersen:2005arxiv} the importance of dimensional reduction is discussed.

In the present work, we examine the $O(2)$ model, with quartic self-coupling at finite density within the 2PI framework. This is the simplest model in which one can study how the chemical potential is implemented in the 2PI formalism. Even though we focus on the so-called two-loop $\Phi$-derivable approximation, we try to describe, as generally as possible, some of the subtleties the introduction of a chemical potential leads to in the 2PI formalism. This work will thus serve as the basis for future applications, in other models. One possible application is the relativistic description of superfluidity, where a Hartree-Fock level 2PI treatment has already been carried out in \cite{Alford:2013koa}. In this work questions were raised concerning renormalization, which were clarified in \cite{GI:2014arxiv}. Another possible application is the question of pion condensation. It has been studied using the 2PI framework in the Hartree-Fock and lowest order $1/N$ truncations, neglecting vacuum fluctuations in \cite{Shu:2007na}. The inclusion of vacuum fluctuations was done in \cite{Andersen:2006ys} using the $O(2N)$ model at the lowest order of the 2PI-$1/N$ expansion. These efforts were further extended in \cite{Bowman:2010thesis}. For more information on pion condensation we point the reader to the functional renormalization group study of \cite{Svanes:2010we} and references therein. The natural step forward would be to consider the next-to-leading order approximation in this $1/N$ expansion. Although doable, this is particularly demanding numerically. A simpler way to consider corrections to the results of \cite{Andersen:2006ys} is to consider the two-loop $\Phi$-derivable approximation, which is also the natural extension for the superfluidity studies.\footnote{One could also consider the Hartree-Fock approximation in the case of the pion condensation. However, a known problem of this approximation is that, in the chiral limit, the order of the phase transition is first order, contrary to what is usually expected.} In this work, we test this approximation in the simpler framework of the $O(2)$ model.

The paper is structured as follows. In Sec.~II we describe the model, the corresponding Euclidean action at finite temperature and finite chemical potential, and its symmetries. In Sec.~III we introduce the 2PI formalism at finite chemical potential. The fact that the Euclidean action is complex at non zero (real) chemical potential leads to certain complications in the 2PI formalism (and in fact in any approach based on a Legendre transform) which we describe and deal with. We also derive the equations in the two-loop $\Phi$-derivable approximation. In Sec.~IV, we relate the Silver Blaze property mentioned above (and its generalization to $n$-point functions) to the transformation property of the Euclidean action under certain gauge transformations of the charged field, and from this we deduce a simple condition for the Silver Blaze property to be fulfilled in any given $\Phi$-derivable approximation. In Sec.~V, we briefly discuss the renormalization of the two-loop $\Phi$-derivable approximation in the presence of a finite chemical potential by making use of the Silver Blaze property. In Sec.~VI we discuss our numerical results for the two-loop approximation together with a qualitative comparison to existing lattice results \cite{Gattringer:2012df}. Various technical details are gathered in the Appendices.

\section{Generalities}\label{sec:gen}

In this work we consider a two-component real scalar field $\varphi=(\varphi_1,\varphi_2)^{\rm t}$ whose Lagrangian density\footnote{A summation over repeated indices is implied.}
\beq
{\cal L}=\frac{1}{2}(\partial^\mu\varphi_a)(\partial_\mu\varphi_a)-\frac{1}{2}m^2_{\rm b}\varphi_a\varphi_a-\frac{\lambda_{\rm b}}{48}(\varphi_a\varphi_a)^2
\eeq
is invariant under $O(2)$ transformations. Any such transformation is either an element of $SO(2)$ or the product of an element of $SO(2)$ by the reflection $(\varphi_1,\varphi_2)\rightarrow (\varphi_1,-\varphi_2)$. It will be sometimes convenient to reformulate the problem in terms of the field $\chi=(\Phi,\Phi^*)^{\rm t}$ with two complex components $\Phi\equiv(\varphi_1+i\varphi_2)/\sqrt{2}$ and $\Phi^*\equiv(\varphi_1-i\varphi_2)/\sqrt{2}$. The Lagrangian density then takes the form
\beq
{\cal L}=\partial^\mu\Phi^*\partial_\mu\Phi-m^2_{\rm b}\Phi^*\Phi-\frac{\lambda_{\rm b}}{12}(\Phi^*\Phi)^2\,,
\eeq
the $SO(2)$ symmetry translates into $U(1)$ symmetry and the reflection symmetry into charge conjugation symmetry $\Phi\leftrightarrow\Phi^*$. Going from the real field formulation in terms of $\phi$ to the complex field formulation in terms of $\chi$ amounts to the change of variables $\chi=U\varphi$ with
\beq\label{eq:U}
U=\frac{1}{\sqrt{2}}\left(\begin{array}{cc}
1 & i\\
1 & -i
\end{array}\right),
\eeq
and we shall make use of this remark whenever it is convenient.

To the continuous and global $SO(2)$ or $U(1)$ invariance, Noether's theorem associates a conserved charge\footnote{We use a different sign convention for the charge than the one found in the standard references \cite{Kapusta:2006pm,LeBellac:1996}. Our convention coincides with that used in \cite{Gattringer:2012df} and is such that a positive chemical potential corresponds to a positive charge density.}
\beq
Q=\int d^3x\,\,[\pi_2\varphi_1-\pi_1\varphi_2]
\eeq
which enters the grand canonical partition function as
\beq\label{eq:Z0}
Z={\rm Tr}\,e^{-\beta(H-\mu Q)}\,,
\eeq
where $\mu$ is the corresponding chemical potential and $\beta=1/T$ is the inverse temperature. It is well known that the partition function (\ref{eq:Z0}) can be given a functional integral representation \cite{Kapusta:2006pm}
\beq\label{eq:Z}
Z\propto\int_{\rm PBC}{\cal D}[\varphi_1,\varphi_2]\,\,e^{-\int_x\,{\cal L}_E}\,,
\eeq
where $\int_x$ stands for $\int_0^{\beta} d\tau\int d^3x$, the functional integration is to be performed over fields obeying the periodic boundary condition (PBC) $\varphi_a(0,\vec{x})=\varphi_a(\beta,\vec{x})$, and the Euclidean Lagrangian density ${\cal L}_E$ is given by
\beq\label{eq:LE1}
{\cal L}_E & = & \frac{1}{2}(\dot\varphi_1+i\mu\varphi_2)^2+\frac{1}{2}(\dot\varphi_2-i\mu\varphi_1)^2\nonumber\\
& & +\,\frac{1}{2}(\nabla\varphi_a)(\nabla\varphi_a)+\frac{1}{2}m^2_{\rm b}\varphi_a\varphi_a+\frac{\lambda_{\rm b}}{48}(\varphi_a\varphi_a)^2\!\,,\nonumber\\
\eeq
or, in the complex field formulation, by
\beq\label{eq:LE2}
{\cal L}_E & = & (\dot\Phi^*-\mu\Phi^*)(\dot\Phi+\mu\Phi)\nonumber\\
& & +\,(\nabla\Phi^*)(\nabla\Phi)+m^2_{\rm b}\Phi^*\Phi+\frac{\lambda_{\rm b}}{12}(\Phi^*\Phi)^2\,.
\eeq
The presence of the chemical potential $\mu$ in ${\cal L}_E$ does not break the $SO(2)$ or $U(1)$ symmetry, as it can be readily checked using Eq.~(\ref{eq:LE1}) or Eq.~(\ref{eq:LE2}).\footnote{In the case of Eq.~(\ref{eq:LE1}), one can rewrite the first two terms as $$\frac{1}{2}\dot\varphi_a\dot\varphi_a-\frac{1}{2}\mu^2\varphi_a\varphi_a+i\mu\left(\begin{array}{ccc}
0\\
0\\
1
\end{array}\right)\cdot\left(\left(
\begin{array}{c}
\dot\varphi_1\\
\dot\varphi_2\\
0
\end{array}
\right)\wedge\left(\begin{array}{c}
\varphi_1\\
\varphi_2\\
0
\end{array}
\right)\right)$$ which makes the invariance under $SO(2)$ manifest.} In contrast, reflection or charge conjugation symmetry is explicitly broken, in agreement with the fact that a non-vanishing chemical potential discriminates between particles and anti-particles. We note also that the presence of the chemical potential does not break parity but breaks ``time-reversal'' defined here as $\Phi(\tau,\vec{x})\to\Phi(\beta-\tau,\vec{x})$ and $\Phi^*(\tau,\vec{x})\to\Phi^*(\beta-\tau,\vec{x})$.

Finally, it is easily checked using Eq.~(\ref{eq:LE2}) that the Lagrangian density, considered as a functional for fields with arbitrary boundary conditions, is invariant under a gauge transformation of the form
\beq\label{eq:gauge}
\Phi\to e^{i\alpha \tau} \Phi\,, \quad \Phi^*\to e^{-i\alpha \tau} \Phi^*\,,
\eeq
provided the external parameter $\mu$ is changed simultaneously according to $\mu\to\mu-i\alpha$. This property plays a role in the general discussion of the Silver Blaze property, to be given in Sec.~\ref{sec:SB}, and has interesting consequences such as the renormalizability of the model at finite $\mu$. We note that, if one wants to maintain the periodicity of the fields under such a transformation, $\alpha$ must be chosen equal to a bosonic Matsubara frequency $\omega_n=(2\pi/\beta)n$, with $n\in\mathds{Z}$.

\section{The 2PI effective action at finite chemical potential}

\subsection{General considerations}\label{sec:2PIgen}
The chemical potential $\mu$ enters the Euclidean Lagrangian densities (\ref{eq:LE1}) and (\ref{eq:LE2}) through their quadratic parts only, whose kernels are denoted respectively $G^{-1}_0$ and ${\cal G}^{-1}_0$ in what follows. It is then straightforward to obtain the 1PI or 2PI effective actions at finite $\mu$, by updating $G_0^{-1}$ or ${\cal G}^{-1}_0$ in the corresponding formulae at $\mu=0$ (there is a subtle point hidden here, related to the fact that the Euclidean action becomes complex at finite $\mu$, but we postpone its discussion until the end of the subsection). For instance, in the real field formulation (\ref{eq:LE1}), the 2PI effective action takes the usual form
\begin{widetext}
\beq\label{eq:2PI0}
\Gamma[\phi,G]=\frac{1}{2}{\rm Tr}\,\big[\ln G^{-1}+G^{-1}_0(G+\phi\phi^{\rm t})\big]+\Gamma_{\rm int}[\phi,G]\,,
\eeq
where `${\rm Tr}$' stands both for an integration over imaginary time and position and for a summation over the internal indices of the field,  while the free inverse propagator reads
\beq\label{eq:G0op}
G^{-1}_0(\tau,\vec{x};\tau',\vec{x}')=\delta(\tau-\tau')\delta^{(3)}(\vec{x}-\vec{x}')\left(
\begin{array}{cc}
-\frac{\partial^2}{\partial{\tau'}^2}-\Delta+m^2_{\rm b}-\mu^2 & -2i\mu\frac{\partial}{\partial\tau'}\\
+2i\mu\frac{\partial}{d\tau'} & -\frac{\partial^2}{\partial{\tau'}^2}-\Delta+m^2_{\rm b}-\mu^2
\end{array}
\right).
\eeq
\end{widetext}
The variable $\phi=\langle\varphi\rangle$ is a vector-valued one-point function that represents the expectation value of the field $\varphi$ in the presence of external local and bilocal sources. The variable $G=\langle\varphi\varphi^{\rm t}\rangle-\langle\varphi\rangle\langle\varphi^{\rm t}\rangle$ is a matrix-valued two-point function that represents the connected correlator of the field $\varphi$ in the presence of the same sources. By construction, it obeys the property
\beq\label{eq:prop}
G_{ab}(x,y)=G_{ba}(y,x)\,.
\eeq
Finally, $-\Gamma_{\rm int}[\phi,G]$ is the sum of two-particle-irreducible (2PI) diagrams that one can draw using the propagator $G$ and the vertices of the shifted theory, defined by the action $S[\phi+\varphi]$.  The standard 1PI effective action is obtained as $\Gamma[\phi]=\Gamma[\phi,\bar G_\phi]$ where $\bar G_\phi$ obeys the stationarity condition $0=\delta\Gamma[\phi,G]/\delta G|_{\phi,\bar G_\phi}$ with the derivative $\delta/\delta G$ taken in the space of propagators obeying (\ref{eq:prop}). Owing to this stationarity condition, the extrema $\bar\phi$ of $\Gamma[\phi]$ can be obtained from the equation $0=\delta\Gamma[\phi, G]/\delta \phi|_{\bar\phi,\bar G_{\bar\phi}}$.

In the case of a homogeneous system, to which we restrict in this work, the field $\bar\phi$ and the propagator $\bar G_{\bar\phi}$ are translation invariant: $\bar\phi(x)=\bar\phi$ and $\bar G_{\bar\phi}(x,y)=\bar G_{\bar\phi}(x-y)$. It is then enough to restrict the 2PI effective action to fields and propagators of this form, in which case one can factor out a trivial volume factor $\beta V$. This defines the 2PI effective potential
\beq\label{eq:2PI1}
\gamma[\phi,G]=\frac{1}{2}\int_Q^T\big[\ln G^{-1}+G^{-1}_0(G+\phi\phi^{\rm t})\big]+\gamma_{\rm int}[\phi,G]\nonumber\\
\eeq
which we have expressed conveniently in terms of the Fourier transform 
\beq
G_{ab}(Q) &= & T\sum_{n=-\infty}^{+\infty} \int\frac{d^3q}{(2\pi)^3}e^{i\omega_n\tau-i\vec{q}\cdot\vec{x}}G_{ab}(\tau,\vec{x}).\nonumber\\
\eeq
Owing to Eq.~(\ref{eq:prop}), it is such that
\beq\label{eq:prop2}
G_{ab}(Q)=G_{ba}(-Q)\,.
\eeq
We have also
\beq
G^{-1}_0(Q)=\left(
\begin{array}{cc}
Q^2-\mu^2+m^2_{\rm b} & -2\mu\omega_n\\
+2\mu\omega_n & Q^2-\mu^2+m^2_{\rm b}
\end{array}
\right),
\eeq
where $Q^2\equiv\omega^2_n+q^2$. The standard 1PI effective potential is obtained as $\gamma(\phi)=\gamma[\phi,\bar G_\phi]$ where $\bar G_\phi$ obeys the stationarity condition $0=\delta\gamma[\phi,G]/\delta G|_{\phi,\bar G_\phi}$, with the derivative $\delta/\delta G$ (which includes a factor $(2\pi)^3/T$) taken in the space of propagators obeying (\ref{eq:prop2}). Accordingly, the extrema $\bar\phi$ of $\gamma(\phi)$ can be obtained from the equation $0=\partial\gamma[\phi,G]/\partial\phi|_{\bar\phi,\bar G_{\bar\phi}}$.\\

A word of caution now. Even though the formulae (\ref{eq:2PI0}) and (\ref{eq:2PI1}) turn out to be correct, it is important to keep in mind that, owing to the presence of the (real) chemical potential, the Euclidean Lagrangian density is complex. So, in general, and in contrast to what occurs at $\smash{\mu=0}$, the Legendre variables that enter the 1PI or 2PI effective actions are not real-valued, even if the original sources are taken real-valued. If we consider for instance the 1PI effective action and we work with the real field formulation (\ref{eq:LE1}), it is natural to introduce a generating functional $W_\mu[J]$ for connected Green functions such that the real-valued field $\varphi$ is coupled to a real-valued source $J$. However, at non-zero $\mu$, the corresponding Legendre variable $\phi$, which is nothing but the expectation value of $\varphi$ in the presence of the source, is usually complex-valued.\footnote{We illustrate this point in App.~\ref{app:complex} where we give the explicit relation between the source and the Legendre variable in the case of the free theory.} Moreover the Legendre variable is not completely unconstrained because it should correspond to a real-valued source through the inverse Legendre transform. One way to circumvent these difficulties is to work from the beginning with a complex-valued source to which corresponds an unconstrained complex-valued Legendre variable. However this does not completely solve the problem because one still needs to show that the observables that one computes within this extended framework are real. 

Fortunately, we do not need to extend the sources in the present work because, as we show in App.~\ref{app:complex}, if the source $J$ is real-valued and homogeneous,\footnote{In fact, we only need to assume that the source is static.} the corresponding Legendre variable $\phi$ is also real-valued and homogeneous. In fact the functional $W_\mu[J]$ itself is real for any real and homogeneous $J$. A similar result holds for the 2PI effective action: for a homogeneous system, both the homogeneous field $\bar\phi$ and the Fourier transform $\bar G_{\bar\phi}(Q)$ of the translationally invariant propagator are real. So, despite the fact that the Euclidean action is complex at finite $\mu$, we can assume that all the variables that enter the 2PI effective potential (\ref{eq:2PI1}) are real. In particular, any thermodynamical observable which is derived from this potential is real.

Yet another difficulty related to the fact that the Euclidean Lagrangian density is complex at finite $\mu$ is that it is not obvious a priori whether one should consider minima or maxima of the effective potential $\gamma(\phi)$. The reason is that in the case of a homogeneous source and for a given non-zero $\mu$, it is not obvious whether the (real-valued) function $W_\mu[J]$ is convex for every $\mu$. Usual proofs of convexity require a positive definite measure. In the present case, it is simple to rewrite the path integral in such a way that the measure is real, but we were not able to ensure that its sign be always positive when $\mu\neq 0$, see App.~\ref{app:complex}. This indicates that continuum approaches based on the use of effective actions may not be able to completely elude the sign problem. We should however qualify this as a ``small sign problem''  in the sense that it does not prevent actual calculations but only makes it difficult to decide which solution to choose. In the present model where the sign problem can be solved using for instance the flux tube representation of the partition function \cite{Gattringer:2012df}, it is probably possible to solve the small sign problem. In the present work, we will not try to do so. For homogeneous $J$, $W_\mu[J]$ is convex for $\mu=0$ and probably also for small $\mu$, see App.~\ref{app:complex}. We shall assume (but we have currently no proof for this) that it is in fact convex for any $\mu$ which implies that one should look for the minimum of $\gamma(\phi)$.

\subsection{Symmetry constraints}
In practice, and following the discussion about translation invariance in the previous section, it is convenient to use as many symmetries of the problem as possible in order to constrain the form of $\bar G_\phi$ (with homogeneous $\phi$) and thus the space of propagators that it is sufficient to restrict to. For instance, in the $O(N)$ model at $\mu=0$, the symmetries impose that $\bar G_\phi$ has only longitudinal and transversal components, that is $\bar G^\phi_{ab}=\bar G_L P^L_{ab}+\bar G_T P^T_{ab}$, with 
\beq
P^L_{ab}\equiv\frac{\phi_a\phi_b}{\phi^2} \quad {\rm and} \quad P^T_{ab}\equiv\delta_{ab}-P^L_{ab}\,.
\eeq This allows one to restrict the 2PI effective potential to propagators admitting the same decomposition, that is to consider the restricted functional $\gamma[\phi,G_L,G_T]\equiv\gamma[\phi,G_L P^L+G_T P^T]$.

In the present $O(2)$ model at finite $\mu$, the symmetries alone do not constrain enough the structure of the propagator (this has to do with the fact that the symmetry group is abelian) and we need some additional input. Using $SO(2)$ invariance together with the explicit form of the 2PI effective action, we show in App.~\ref{app:tensor} that
\beq\label{eq:decomp}
\bar G^\phi_{ab}=\bar G_LP^L_{ab}+\bar G_TP^T_{ab}+\bar G_A\,\varepsilon_{ab}\,,
\eeq
with $\varepsilon_{ab}$ the antisymmetric tensor such that $\varepsilon_{12}=1$ and where, owing to (\ref{eq:prop2}), $\bar G_{L,T}(-Q)=\bar G_{L,T}(Q)$ and $\bar G_A(-Q)=-\bar G_A(Q)$. More precisely, because parity is manifest, $G_{L,T,A}(Q)$ are invariant under $\vec{q}\to -\vec{q}$ whereas, in contrast, ``time reversal'' is broken by the presence of the chemical potential, which is reflected in the fact that $G_A(Q)$ is not invariant, but rather changes sign, under $\omega_n\to -\omega_n$. The presence of the skew-component $\bar G_A$ stems from the fact that, in contradistinction with the case $\smash{\mu=0}$, we do not have $\bar G^\phi_{ab}(Q)=\bar G^\phi_{ba}(Q)$, see App.~\ref{app:tensor}. 

The decomposition (\ref{eq:decomp}) shows that it is enough to consider the restricted functional $\gamma[\phi,G_L,G_T,G_A]\equiv\gamma[\phi,G_LP^L+G_TP^T+G_A\,\varepsilon]$. The usual 1PI effective potential is obtained as $\gamma(\phi)=\gamma[\phi,\bar G_L,\bar G_T,\bar G_A]$ where the propagators $\bar G_{L,T,A}$ are determined from the stationarity conditions or {\it gap equations}
\beq\label{eq:gapu}
0=\left.\frac{\delta\gamma[\phi,G_L,G_T,G_A]}{\delta G_{L,T,A}}\right|_{\phi,\bar G_L,\bar G_T,\bar G_A}\,,
\eeq
where $\delta/\delta G_{L,T,A}$ are derivatives in the space of propagators such that $G_{L,T}(-Q)=G_{L,T}(Q)$ and $G_A(-Q)=-G_A(Q)$. The physical value of the one-point function is obtained at the minimum of $\gamma(\phi)$ which, as any other extremum $\bar\phi$, and owing to the stationarity conditions (\ref{eq:gapu}), obeys the {\it field equation}
\beq\label{eq:fieldu}
0=\left.\frac{\delta\gamma[\phi,G_L,G_T,G_A]}{\delta\phi_a}\right|_{\bar\phi,\bar G_L,\bar G_T,\bar G_A}\,,
\eeq
\vglue2mm
\noindent{where it is understood that $\bar G_{L,T,A}$ are evaluated at $\phi=\bar\phi$. In what follows we shall solve the gap and field equations (\ref{eq:gapu})-(\ref{eq:fieldu}) in the two-loop $\Phi$-derivable approximation, to be defined in the next section.}

\subsection{The two-loop approximation}\label{sec:2loop}
The 2PI effective action cannot be computed exactly and some approximation is required. Here, we consider the two-loop $\Phi$-derivable approximation which amounts to keeping in $\Gamma_{\rm int}[\phi,G]$ diagrams up to two-loop order. After rescaling the field and the propagator as $\phi\to\sqrt{Z_2}\phi$ and $G\to Z_0G$, the 2PI effective action to this order of approximation reads
\begin{widetext}
\beq\label{eq:2PI}
\Gamma[\phi,G] & = & \frac{1}{2}\int_x{\rm tr}\,\big[\ln G^{-1}+G^{-1}_0\,(Z_0G+Z_2\phi\phi^{\rm t})\big](x,x)+\frac{\lambda_4}{48}\,\int_x(\phi^{\rm t}(x)\phi(x))^2\nonumber\\
& & +\,\frac{\lambda_2^{(A)}}{24}\,\int_x\phi^{\rm t}(x)\phi(x)\,{\rm tr}\,G(x,x)+\frac{\lambda_2^{(B)}}{12}\,\int_x\phi^{\rm t}(x)G(x,x)\phi(x)+\frac{\lambda_0^{(A)}}{48}\,\int_x \big[{\rm tr}\,G(x,x)\big]^2+\frac{\lambda_0^{(B)}}{24}\,\int_x {\rm tr}\,G^2(x,x)\nonumber\\
& & -\,\frac{\lambda^2_\star}{144}\int_x\int_y\phi^{\rm t}(x)G(x,y)\phi(y)\,{\rm tr}\,\big[G(x,y)G(y,x)\big]-\frac{\lambda^2_\star}{72}\int_x\int_y\phi^{\rm t}(x)G(x,y)G(y,x)G(x,y)\phi(y)\,,
\eeq
where we have dropped an infinite term proportional to $\beta V$. The need for two field-strength renormalization factors $Z_0$ and $Z_2$ and five bare couplings $\lambda_0^{(A,B)}$, $\lambda_2^{(A,B)}$ and $\lambda_4$ will be discussed in Sec.~\ref{sec:renormalization}. We shall also need to introduce two different bare masses $Z_0m^2_{\rm b}\to m^2_0$ and $Z_2 m^2_{\rm b}\to m^2_2$. All these bare parameters will be fixed in terms of two renormalized parameters $m_\star$ and $\lambda_\star$, as it should because there are only two free parameters in the model.

Restricting the 2PI effective action to translationally invariant fields and propagators and to propagators that admit the decomposition (\ref{eq:decomp}), one obtains the restricted 2PI effective potential
\beq\label{eq:2PI_LTA}
& & \gamma[\phi,G_L,G_T,G_A]\nonumber\\
& & \hspace{0.5cm}=\,\frac{1}{2}\int_Q^T \big[-\ln (G_L(Q)G_T(Q)+G^2_A(Q))+(Z_0(Q^2-\mu^2)+m^2_0)\,(G_L(Q)+G_T(Q))+4Z_0\mu\omega_n G_A(Q)\big]\nonumber\\
& & \hspace{0.5cm}+\Bigg(m_2^2-\mu^2 Z_2+\,\frac{\lambda_4}{48}\phi^2+\frac{\lambda^{(A+2B)}_2}{24}{\cal T}[G_L]+ \frac{\lambda^{(A)}_2}{24}{\cal T}[G_T]\Bigg)\,\phi^2+\frac{\lambda^{(A+2B)}_0}{48}\,\left({\cal T}^2[G_L]+{\cal T}^2[G_T]\right)+\frac{\lambda^{(A)}_0}{24}{\cal T}[G_L]{\cal T}[G_T]\nonumber\\
& & \hspace{0.5cm}-\,\frac{\lambda^2_\star}{144}\phi^2(3{\cal S}[G_L]+{\cal S}[G_L;G_T;G_T])-\frac{\lambda^2_\star}{72}\phi^2(3{\cal S}[G_A;G_A;G_L]-{\cal S}[G_A;G_A;G_T])\,,
\eeq
\end{widetext}
where we have introduced similar notations as in \cite{Marko:2013lxa}, namely $\lambda^{(\alpha A+\beta B)}_{0,2}\equiv\alpha\lambda^{(A)}_{0,2}+\beta\lambda^{(B)}_{0,2}$ and
\begin{subequations}
\beq
{\cal T}[G] & \equiv & \int_Q^T G(Q)\,,\\
{\cal B}[G_1;G_2](K) & \equiv & \int_Q^T G_1(Q)G_2(Q+K)\,,\\
{\cal S}[G_1;G_2;G_3] & \equiv & \int_Q^T\int_K^T G_1(Q)G_2(K)G_3(Q+K)\,,\nonumber\\
\eeq
\end{subequations}
as well as ${\cal B}[G](K)\equiv{\cal B}[G;G](K)$ and ${\cal S}[G]\equiv{\cal S}[G;G;G]$ (the function ${\cal B}$ is needed below).  It is to be noticed that, in the case where all propagators are even, the relative sign between $Q$ and $K$ in the definitions of ${\cal B}$ and ${\cal S}$ can be chosen arbitrarily, but in all the other cases some care needs to be taken.

Because of SO(2) invariance, the restriction of (\ref{eq:2PI}) to translationally invariant fields and propagators and to propagators that admit the decomposition (\ref{eq:decomp}) depends on the vector $\phi$ only through $\phi^2$. Then, in order to obtain (\ref{eq:2PI_LTA}), we chose conveniently  $\phi_a=\sqrt{\phi^2}\delta_{a1}$, in which case we have $P^L_{ab}=\delta_{a1}\delta_{b1}$ and $P^T_{ab}=\delta_{a2}\delta_{b2}$, from which it follows that
\beq\label{eq:gg}
G=\left(
\begin{array}{cc}
G_L & G_A\\
-G_A & G_T
\end{array}
\right)
\eeq
and thus for instance ${\rm tr}\ln G=\ln {\rm det}\,G=\ln (G_LG_T+G_A^2)$. The formula (\ref{eq:2PI_LTA}) is of course valid for any direction $\phi_a$.

The gap equations (\ref{eq:gapu}) are equivalent to
\beq
\frac{\bar G_{L,T}(K)}{\bar G_L(K)\bar G_T(K)+\bar G^2_A(K)} & = & K^2-\mu^2+\bar M^2_{T,L}(K)\,,\nonumber\\
\frac{\bar G_A(K)}{\bar G_L(K)\bar G_T(K)+\bar G^2_A(K)} & = & 2\mu\omega_n-\bar M^2_A(K)\,,\label{eq:aA}
\eeq
with $K^2=\omega^2_n+k^2$ and
\begin{subequations}
\beq
\bar M^2_L(K) & = & (K^2-\mu^2)(Z_0-1)+m^2_0\nonumber\\
& + & \frac{\lambda^{(A+2B)}_0}{12}{\cal T}[\bar G_L]+\frac{\lambda^{(A)}_0}{12}{\cal T}[\bar G_T]\nonumber\\
& + & \frac{\phi^2}{12}\left[\lambda^{(A+2B)}_2-\frac{3\lambda^2_\star}{2}{\cal B}[\bar G_L](K)\right.\nonumber\\
& & \hspace{0.8cm}\left.-\,\frac{\lambda^2_\star}{6}{\cal B}[\bar G_T](K)+\lambda^2_\star{\cal B}[\bar G_A](K)\right]\!,\label{eq:ML}\\
\bar M^2_T(K) & = & (K^2-\mu^2)(Z_0-1)+m^2_0\nonumber\\
& + & \frac{\lambda^{(A)}_0}{12}{\cal T}[\bar G_L]+\frac{\lambda^{(A+2B)}_0}{12}{\cal T}[\bar G_T]\nonumber\\
&  + & \frac{\phi^2}{12}\left[\lambda^{(A)}_2-\frac{\lambda^2_\star}{3}{\cal B}[\bar G_L;\bar G_T](K)\right.\nonumber\\
&  & \hspace{3.0cm}\left.-\,\frac{\lambda^2_\star}{3}{\cal B}[\bar G_A](K)\big]\right]\!,\label{eq:MT}\\
\bar M^2_A(K) & = & -2(Z_0-1)\mu\omega_n\nonumber\\
& + & \frac{\lambda^2_\star}{12}\phi^2\left[\frac{1}{3}{\cal B}[\bar G_T;\bar G_A](K)-{\cal B}[\bar G_L;\bar G_A](K)\right]\!.\label{eq:MA}\nonumber\\
\eeq
\end{subequations}
The propagator components $\bar G_{L,T,A}$ can be expressed in terms of the momentum dependent gap masses $\bar M^2_{L,T,A}$ by inverting the relations (\ref{eq:aA}). One obtains
\begin{subequations}
\beq
\bar G_{L,T}(K) & = & (K^2-\mu^2+\bar M^2_{T,L}(K))\Delta^{-1}(K)\,,\label{eq:GT}\\
\bar G_A(K) & = & (2\mu\omega_n-\bar M^2_A(K))\Delta^{-1}(K)\,,\label{eq:GA}
\eeq
with
\beq
\Delta(K) & = & \prod_{i=L,T}\big(K^2-\mu^2+\bar M^2_i(K)\big)\nonumber\\
& + & \big(2\mu\omega_n-\bar M^2_A(K)\big)^2\,.
\eeq
\end{subequations}
Finally, the field equation (\ref{eq:fieldu}) takes the form
\beq
0 & = & \bar\phi_a\left(-\mu^2 Z_2+m^2_2+\frac{\lambda_4}{12}\phi^2\right.\nonumber\\
& & +\,\frac{\lambda^{(A+2B)}_2}{12}{\cal T}[\bar G_L]+\frac{\lambda^{(A)}_2}{12}{\cal T}[\bar G_T]\nonumber\\
& & -\,\frac{\lambda^2_\star}{24}{\cal S}[\bar G_L]-\frac{\lambda^2_\star}{72}{\cal S}[\bar G_L;\bar G_T;\bar G_T]\nonumber\\
& & \left.-\,\frac{\lambda^2_\star}{12}{\cal S}[\bar G_L;\bar G_A;\bar G_A]+\frac{\lambda^2_\star}{36}{\cal S}[\bar G_T;\bar G_A;\bar G_A]\right).\nonumber\\
\label{eq:field}
\eeq
We obtain the numerical solution presented in Sec.~\ref{sec:results} by iterating the coupled gap and
field equations. We also compute derived quantities like the density or the curvature masses, which
depend on the solutions, using the same routines. More details on our numerical method can be
found in Sec.~\ref{sec:results}.

\subsection{Phase transition}\label{sec:transition}

In order to study the phase transition, we shall monitor the nature of the extrema $\bar\phi$ that solve the field equation by computing the curvature tensor $\partial^2\gamma/\partial\phi_a\partial\phi_b$ at $\phi=\bar\phi$. We note that the $SO(2)$ invariance of $\gamma(\phi)$ implies that $\gamma(\phi)=g(\phi^2)$ and thus the curvature tensor has the general structure
\begin{subequations}
\beq
\frac{\partial^2\gamma}{\partial\phi_a\partial\phi_b}=\gamma^{(2)}_LP^L_{ab}+\gamma^{(2)}_TP^T_{ab}
\eeq
with
\beq
\gamma^{(2)}_L & = & 2g'(\phi^2)+4\phi^2 g''(\phi^2)\,,\\
\gamma^{(2)}_T & = & 2g'(\phi^2)\,.
\eeq
\end{subequations}
Because the field equation takes the form $0=g'(\bar\phi^2)\bar\phi_a$, we have two types of solutions. Those for which $\bar\phi=0$, in which case $\gamma^{(2)}_L=\gamma^{(2)}_T=2g'(0)$ and those for which $g'(\bar\phi^2)=0$ in which case $\gamma^{(2)}_L=4\bar\phi^2g''(\bar\phi^2)$ and $\gamma^{(2)}_T=0$. This last identity is nothing but the Goldstone theorem. 

In what follows, we find it more convenient to work with the {\it curvature mass} tensor
\beq
\hat M^2_{ab}\equiv\frac{\partial^2\gamma}{\partial\phi_a\partial\phi_b}+\delta_{ab}\mu^2=\hat M^2_LP^L_{ab}+\hat M^2_TP^T_{ab}\,,
\eeq
with $\hat M^2_{L,T}=\gamma^{(2)}_{L,T}+\mu^2$. The reason for considering this tensor is that, in the exact theory, it coincides with the {\it gap mass} tensor $\bar M^2_{ab}(K=0)\equiv\bar M^2_{ab}$. In particular, in the exact theory and in the broken phase, both $\hat M^2_T$ and $\bar M^2_T$ obey Goldstone theorem in the form $\hat M^2_T=\bar M^2_T=\mu^2$, just as in the general discussion of \cite{Watanabe:2013uya}. In a given truncation of the 2PI effective action, such as the two-loop truncation considered here, $\bar M^2_T$ generically violates the Goldstone theorem, even though, as we will see, it could be almost satisfied in certain regions of the $(\mu,T)$ plane.\\

Coming back to the discussion of the various solutions of the field equation, we note that in order for $\bar\phi=0$ to be considered the physical solution, that is corresponding to the absolute minimum of $\gamma(\phi)$, a necessary condition is that $g'(0)\geq 0$. In the case where the transition is second order, as we will find it to be in the present approximation, one moves continuously from a symmetric phase solution $\bar\phi=0$ to a broken phase solution $\bar\phi\neq 0$. The {\it transition line} $\mu_{\rm c}(T)$ or $T_{\rm c}(\mu)$ in the $(\mu,T)$ plane where this occurs is determined from the condition $g'(0)=0$ or equivalently from the condition that the curvature mass at the origin of the potential $\hat M^2_{\phi=0}=2g'(0)+\mu^2$ becomes equal to $\mu^2$. This mass can be obtained by noting that $g'(\phi)$ can be read off from the expression inside the bracket in the r.h.s. of Eq.~(\ref{eq:field}) and by using that, when $\phi=0$, $\bar G^{\phi=0}_L=\bar G^{\phi=0}_T\equiv\bar G_I$. One gets then
\beq\label{eq:curv1}
\hat M^2_{\phi=0} & = & -\mu^2 (Z_2-1)+m^2_2+\frac{\lambda^{(A+B)}_2}{6}{\cal T}[\bar G_I]\nonumber\\
& & -\,\frac{\lambda^2_\star}{18}\Big({\cal S}[\bar G_I]+{\cal S}[\bar G_I;\bar G^{\phi=0}_A;\bar G^{\phi=0}_A]\Big).
\eeq
Moreover, since at $\phi=0$, $\bar M^2_{A,\phi=0}(K)=-2(Z_0-1)\mu\omega_n$ and $\bar M^2_{L,\phi=0}(K)=\bar
M^2_{T,\phi=0}(K)\equiv(K^2-\mu^2)(Z_0-1)+\Delta\bar M^2_{\phi=0}$, we have, owing to (\ref{eq:GT}) and (\ref{eq:GA}),
\begin{subequations}
\beq
\bar G_I(K) & = & \frac{Z_0(K^2-\mu^2)+\Delta\bar M^2_{\phi=0}}{(Z_0(K^2-\mu^2)+\Delta\bar
M^2_{\phi=0})^2+4Z_0^2\mu^2\omega^2_n}\,,\nonumber\\\\
\bar G_A^{\phi=0}(K) & = & \frac{2Z_0\mu\omega_n}{(Z_0(K^2-\mu^2)+\Delta\bar
M^2_{\phi=0})^2+4Z_0^2\mu^2\omega^2_n}\,,\nonumber\\
\eeq
\end{subequations}
with 
\beq\label{eq:tad1}
\Delta\bar M^2_{\phi=0}=m^2_0+\frac{\lambda_0^{(A+B)}}{6}{\cal T}[\bar G_I]\,,
\eeq
which follows from either (\ref{eq:ML}) or (\ref{eq:MT}). 

\subsection{Complex formulation}
Before closing this section, let us finally remind that one can always switch to the formulation in terms of the field $\chi=(\Phi,\Phi^*)^{\rm t}$. In fact we could have derived the 2PI effective action directly within this formulation. However, it is simpler to do so indirectly, using the real field formulation and the change of variables (\ref{eq:U}). Using the notations ${\cal X}=\langle\chi\rangle$ and ${\cal G}=\langle\Phi\Phi^\dagger\rangle-\langle\Phi\rangle\langle\Phi^\dagger\rangle$ to denote the expectation value of the field $\chi$ and the corresponding connected correlator, one has ${\cal X}=U\phi$ and ${\cal G}=UGU^\dagger$ with $U$ the matrix given in (\ref{eq:U}). The 2PI effective action in the complex formulation is thus obtained as $\tilde\Gamma[{\cal X},{\cal G}]=\Gamma[\phi,G]$. One gets
\begin{widetext}
\beq\label{eq:2PI3}
\tilde\Gamma[{\cal X},{\cal G}]=\frac{1}{2}{\rm Tr}\,\big[\ln {\cal G}^{-1}+{\cal G}^{-1}_0({\cal G}+{\cal X}\tilde{\cal X})\big]+\tilde\Gamma_{\rm int}[{\cal X},{\cal G}],
\eeq
with $\tilde\Gamma_{\rm int}[{\cal X},{\cal G}]=\Gamma_{\rm int}[\phi,G]$ and
\beq\label{eq:calG0}
{\cal G}_0^{-1}(\tau,\vec{x};\tau',\vec{x}')=\delta(\tau-\tau')\delta^{(3)}(\vec{x}-\vec{x}')\left(
\begin{array}{cc}
-\left(\frac{\partial}{\partial\tau'}+\mu\right)^2-\Delta +m^2_{\rm b} & 0\\
0 & -\left(\frac{\partial}{\partial\tau'}-\mu\right)^2-\Delta +m^2_{\rm b}
\end{array}
\right),
\eeq
\end{widetext}
the kernel of the quadratic form (\ref{eq:LE2}). The same difficulty as in the real field formulation occurs: the fields $\Phi$ and $\Phi^*$ (the components of $\chi$) that enter the path integral being complex conjugate of each other, it is natural to couple them to sources $J^*$ and $J$ which are also complex conjugate of each other. However, the corresponding Legendre variables (the components of ${\cal X}$) will not be complex conjugate of each other in general and we shall rather denote them by ${\cal F}$ and $\bar{\cal F}$ respectively. Thus $\chi=(\Phi,\Phi^*)^{\rm t}$ and ${\cal X}=({\cal F},\bar{\cal F})^{\rm t}$ with ${\cal F}^*\neq\bar{\cal F}$ in general. The notation $\tilde{\cal X}$ in (\ref{eq:2PI3}) stands for $(\bar{\cal F},{\cal F})$. It boils down to ${\cal X}^\dagger$ when $\bar{\cal F}={\cal F}^*$. To check that this is the correct notation to be introduced, we write
\beq
{\rm Tr}\,G^{-1}_0\phi\phi^{\rm t}={\rm Tr}\,U^\dagger {\cal G}^{-1}_0UU^\dagger{\cal X}{\cal X}^{\rm t} U^*={\rm Tr}\,{\cal G}^{-1}_0{\cal X}\tilde{\cal X}\,,\nonumber\\
\eeq
where we have used the cyclicity of the trace and 
\beq
U^*U^\dagger=\left(\begin{array}{cc}
0 & 1\\
1 & 0
\end{array}\right).
\eeq
Again, one can show that, in the case of homogeneous external sources, the components of the expectation value ${\cal X}$ are complex conjugate to each other, that is ${\cal F}^*=\bar{\cal F}$. In this case also $\tilde{\cal X}$ coincides with ${\cal X}^\dagger$. Finally, the standard 1PI effective action is obtained from (\ref{eq:2PI3}) as $\tilde\Gamma[{\cal X}]=\Gamma[{\cal X},\bar{\cal G}_{\cal X}]$ where $\bar{\cal G}_{{\cal X}}$ obeys the stationarity condition $0=\delta\tilde\Gamma/\delta{\cal G}|_{\bar {\cal G}_{\cal X}}$.

The complex field formulation is particularly useful in the symmetric phase. Indeed, owing to $U(1)$ invariance, $\bar{\cal G}_{{\cal X}=0}$ is invariant under $U(1)$ transformations and thus its charged, off-diagonal components, need to vanish, just as for the free propagator derived from (\ref{eq:calG0}). So, in the symmetric phase, it is definitely simpler to work in the complex field formulation because the propagator is diagonal.\footnote{In the broken phase, the propagators have four non-vanishing components in both formulations. There is however a preference for the real field formulation since the four components of the propagator in Fourier space are real, which makes the numerical implementation easier.} A simple application of this remark is the determination of the curvature mass at $\phi=0$. In the form (\ref{eq:curv1}), it is cumbersome (although doable) to perform the Matsubara sums. Instead, we can switch to the complex field formulation, for which the propagator reads $\bar {\cal G}_{{\cal X}=0}=U(\bar G_I\mathds{1}+\bar G_A^{\phi=0} \varepsilon)
U^\dagger=\bar G_I\mathds{1}-i\bar G_A^{\phi=0} \sigma_3$ where we have used that $U\varepsilon U^\dagger=-i\sigma_3$ with $\sigma_3$ the third Pauli matrix. Then
\beq
\bar {\cal G}_{{\cal X}=0}(K)=\left(
\begin{array}{cc}
\bar {\cal D}(K) & 0\\
0 & \bar {\cal D}^*(K)
\end{array}
\right)
\eeq
with
\beq\label{eq:D}
\bar {\cal D}(K)&=&\bar G_I(K)-i\bar G^{\phi=0}_A(K)\nonumber\\
&=&\frac{1}{Z_0((\omega_n+i\mu)^2+q^2)+\Delta\bar M^2_{{\cal X}=0}}\,,
\eeq
where $\Delta\bar M^2_{{\cal X}=0}\equiv\Delta\bar M^2_{\phi=0}$, that is after applying the change of basis to \eqref{eq:tad1}
\beq\label{eq:M20}
\Delta\bar M^2_{{\cal X}=0}=m^2_0+\frac{\lambda_0^{(A+B)}}{6}{\cal T}[\bar {\cal D}]\,.
\eeq
To rewrite the curvature mass $\hat M^2_{{\cal X}=0}$ in terms of $\bar {\cal D}$, it is useful to note that, $\bar G_A$ being odd, integrals such as ${\cal T}[\bar G_A]$ and ${\cal S}[\bar G_A;\bar G_I;\bar G_I]$ vanish. It is then a very simple exercise to show that
\beq\label{eq:curv2}
\hat M^2_{{\cal X}=0} & = & -\mu^2(Z_2-1)+m^2_2+\frac{\lambda_2^{(A+B)}}{6}{\cal T}[\bar {\cal D}]\nonumber\\
&&-\frac{\lambda^2_\star}{18}{\cal S}[\bar {\cal D},\bar {\cal D}^*,\bar {\cal D}]\,,
\label{eq:symm_curvature}
\eeq
The Matsubara sums in (\ref{eq:curv2}) are now easily computed, see App.~\ref{app:sum}, because $\mu$ enters as a mere (imaginary) shift of the Matsubara frequencies.

\section{Silver Blaze property}\label{sec:SB}
 Before discussing our results in the two-loop $\Phi$-derivable approximation, we need to explain how the corresponding equations are renormalized, see Sec.~\ref{sec:renormalization}. Our renormalization can be carried out using counterterms which do not depend on the chemical potential (and neither on the temperature). That this is possible can be seen as the consequence of a basic property, known as the Silver Blaze property, and its extension to $n$-point functions. These properties can be seen as consequences of the particular transformation property of the Euclidean Lagrangian density under (\ref{eq:gauge}) which we discussed in Sec.~\ref{sec:gen}, as we now show argue.
 
\subsection{Generalities}

Let us denote by $Z_\mu$ the partition function of the system in the presence of a complex chemical potential $\mu$. As long as the system is in the symmetric phase we expect $Z_\mu$ to be analytic in some strip $|{\rm Re}\,\mu|<\mu_{\rm c}(T)$ with $\mu_c(T)>0$. Now, if we consider a gauge transformation (\ref{eq:gauge}) with $\alpha=\omega_n$, combined with a shift of $\mu$ by $\delta\mu=-i\omega_n$, the invariance of the Lagrangian density under such a transformation and the fact that the boundary conditions on the field remain unchanged imply that the partition function is periodic in $\mu$ with period $i\omega_n$: $Z_{\mu-i\omega_n}=Z_\mu$. At zero temperature,\footnote{More precisely, in the limit $n\to\infty$ with $T=\omega/(2\pi n)$.} this periodicity property translates into the independence of $Z_\mu$ with respect to the imaginary part of $\mu$, and thus, after analytic continuation, to the $\mu$-independence of $Z_\mu$, and in turn of any thermodynamical observable derived from it, in the whole strip $|{\rm Re}\,\mu|<\mu_{\rm c}\equiv\mu_{\rm c}(T=0)$. This is the so-called Silver Blaze property.

The Silver Blaze property can be extended to $n$-point functions in the following way. Using the complex field formulation (\ref{eq:LE2}), let us introduce the generating functional
\beq
e^{W_\mu[J,J^*]}\equiv\int {\cal D}[\Phi,\Phi^*]\,e^{-\int_x {\cal L}_E+\int_x (J^*(x)\Phi(x)+J(x)\Phi^*(x))}\,,\nonumber\\
\eeq
for any complex $\mu$ in the strip $|{\rm Re}\,\mu|<\mu_{\rm c}(T)$. The same argument as above leads to the identity
\beq
W_{\mu-i\omega_n}[e^{i\omega_n \tau} J,e^{-i\omega_n \tau} J^*]=W_\mu[J,J^*]\,.
\eeq
If we denote by $\Gamma_\mu[{\cal F},\bar{\cal F}]$ the Legendre transform  of $W_\mu[J,J^*]$, this result takes the form
\beq
\Gamma_{\mu-i\omega_n}[e^{i\omega_n \tau}{\cal F},e^{-i\omega_n \tau}\bar{\cal F}]=\Gamma_\mu[{\cal F},\bar{\cal F}].
\eeq
Functional derivatives of this identity evaluated at ${\cal F}=\bar{\cal F}=0$ yield the following identities for $2m$-point vertex functions, if the system is in the symmetric phase,
\beq
& & e^{i\omega_n(\tau_1+\dots+\tau_n-\sigma_1-\dots-\sigma_n)}\Gamma^{(m;m)}_{\mu-i\omega_n}(x_1,\dots,x_n;y_1,\dots,y_n)\nonumber\\
& & \hspace{0.5cm}=\,\Gamma^{(m;m)}_\mu(x_1,\dots,x_n;y_1,\dots,y_n)\,.
\eeq
At finite temperature, we do not know how to analytically continue this relation from $i\omega_n$ to $z$ because the continuation is not unique. In contrast, at zero temperature and for $|{\rm Re}\,\mu|<\mu_{\rm c}$, the continuation is unique in the strip $|{\rm Re}(\mu-z)|<\mu_{\rm c}$ and we have
\beq
& & e^{z(\tau_1+\dots+\tau_n-\sigma_1-\dots-\sigma_n)}\Gamma^{(m;m)}_{\mu-z}(x_1,\dots,x_n;y_1,\dots,y_n)\nonumber\\
& & \hspace{0.5cm}=\,\Gamma^{(m;m)}_\mu(x_1,\dots,x_n;y_1,\dots,y_n)\,.
\eeq
In particular, for $z=\mu$ and $|{\rm Re}\,\mu|<\mu_{\rm c}$, we obtain
\beq\label{eq:id}
& & e^{\mu(\tau_1+\dots+\tau_n-\sigma_1-\dots-\sigma_n)}\Gamma^{(m;m)}_0(x_1,\dots,x_n;y_1,\dots,y_n)\nonumber\\
& & \hspace{0.5cm}=\,\Gamma^{(m;m)}_\mu(x_1,\dots,x_n;y_1,\dots,y_n)\,,
\eeq
which shows that, at zero temperature and in the symmetric phase, the dependence of the vertex functions with respect to $\mu$ is trivial and amounts to appropriate phase multiplications in configuration space. Note that for $m=0$, Eq.~(\ref{eq:id}) is nothing but the expression of the Silver Blaze property $Z_\mu=Z_0$ in terms of the Legendre transform $\Gamma_\mu$. For $m>0$, we can go to Fourier space to obtain the following simple generalization of the Silver Blaze property: the $n$-point vertex functions for $T=0$ and  $\mu<\mu_c$, are obtained from those at $T=0$ and $\mu=0$ after shifting the external (continuous) Matsubara frequencies according to $i\omega\to i\omega \pm\mu$ with the sign $\pm$ depending on wether the external leg corresponds to a particle or an anti-particle. We shall see explicit realizations of this property in App.~\ref{app:sum}.

One consequence of the above result is that, at zero temperature and in the symmetric phase, the ultraviolet divergences at finite $\mu$ are exactly those at $\mu=0$. In particular $\mu$ does not require any renormalization factor, in line with similar arguments found in the literature \cite{Benson:1991nj}.  This result is expected since, even though the chemical potential appears effectively as an internal microscopic parameter in Eqs.~(\ref{eq:LE1}) and (\ref{eq:LE2}), it is an external macroscopic parameter, just like the temperature, and as such should not be renormalized. We stress however that the previous argument is valid only in the symmetric phase and at $\smash{T=0}$. Its extension to the broken-phase and/or at finite temperature is given in App.~\ref{app:renormalization}.

\subsection{Silver Blaze and $\Phi$-derivable approximations}
Based on the previous discussion, we expect the Silver Blaze property to be obeyed in a given framework, if the approximation, the UV regularization and the discretisation that one uses, all preserve the transformation property of the Euclidean Lagrangian density under the gauge transformation (\ref{eq:gauge}). For instance, if one has in mind a lattice approach where the imaginary time $\tau$ is discretised according to $\tau_k=k\beta/N\equiv ka$, it is convenient to rewrite the continuum Euclidean Lagrangian density (\ref{eq:LE2}) as
\beq
{\cal L}_E & = & U_{\mu}\partial_\tau(U_{-\mu}\Phi^*) U_{-\mu}\partial_\tau(U_{\mu}\Phi)\nonumber\\
& & +\,(\nabla\Phi^*)(\nabla\Phi)+m^2_{\rm b}\Phi^*\Phi+\frac{\lambda_{\rm b}}{12}(\Phi^*\Phi)^2\,,\label{eq:LE3}
\eeq
with $U_\mu(\tau)=e^{\mu \tau}$ and then to discretise the time derivatives according to
\begin{subequations}
\beq
U_{-\mu}\partial_\tau(U_{\mu}\Phi) & \to & \frac{1}{a}\Big[e^{\mu a}\Phi_{k+1}-\Phi_k\Big]\,,\\
U_{\mu}\partial_\tau(U_{-\mu}\Phi^*) & \to & \frac{1}{a}\Big[e^{-\mu a}\Phi^*_{k+1}-\Phi^*_k\Big]\,.
\eeq
\end{subequations}
The discretised action is then invariant under the change of variables $\Phi_k\to e^{i\omega_n \tau_k}\Phi_k$, $\Phi_k^*\to e^{-i\omega_n \tau_k}\Phi_k^*$, $\mu\to\mu-i\omega_n$ and the Silver Blaze property holds. Introducing the chemical potential through link variables in analogy with the gauge fields is the standard discretization on the lattice, and as shown in \cite{Hasenfratz:1983ba} this avoids the appearance of quadratic divergences in the case of free fermions.

\begin{figure}[t]
\includegraphics[width=0.45\textwidth]{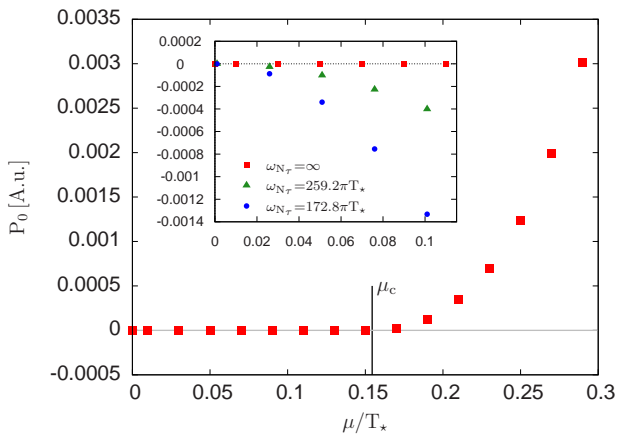}
\caption{Illustration of the Silver Blaze phenomenon based on the differential pressure $P(T,\mu)-P(0,0)$ at $T=0$ as a function of the chemical potential in the BEC case, obtained in our two-loop approximation with parameters: $m_\star^2/T_\star^2=0.1,\,\lambda_\star=3$. In the symmetric phase $\mu<\mu_{\rm c}\equiv\mu_{\rm c}(T=0)$, the pressure is constant, in agreement with the Silver Blaze property. The insert shows violations of the Silver Blaze property which occur if we do not ensure that the largest Matsubara frequency $\omega_{N_\tau}\equiv 2\pi N_\tau T$ goes to infinity as $T\to 0$ (see text for explanation).\label{fig:sb}}
\end{figure}

Let us now discuss the Silver Blaze property within $\Phi$-derivable approximations. To this purpose, it is convenient to employ the complex field formulation (\ref{eq:2PI3}). Because the chemical potential in (\ref{eq:calG0}) is combined with time derivatives in the form of covariant derivatives, it is clear that\footnote{In the real field formulation, this property would read $\int_0^\beta d\tau'G^{-1}_{0,\mu-i\alpha}(\tau,\tau') R(\tau')\phi(\tau')=R(\tau)G^{-1}_{0,\mu}\phi$, with $R(\tau)$ a $SO(2)$ rotation with angle $\alpha\tau$.}
\beq
\int_0^\beta d\tau'{\cal G}^{-1}_{0,\mu-i\alpha}(\tau,\tau')(e^{i\alpha\tau' Q}{\cal X}(\tau'))=e^{i\alpha Q}{\cal G}^{-1}_{0,\mu}{\cal X}(\tau)\nonumber\\
\eeq
where $Q$ is the charge operator and we have made the $\mu$ dependence of ${\cal G}^{-1}_{0,\mu}$ explicit in the notation. It follows that the explicit trace in (\ref{eq:2PI3}) is invariant under ${\cal X}\to e^{i\alpha\tau Q}{\cal X}$, ${\cal G}\to e^{i\alpha\tau Q}{\cal G}e^{-i\alpha\tau' Q}$, provided the chemical potential is changed according to $\mu\to\mu-i\alpha$. Moreover, it is simple to convince oneself that any diagram contributing to $\tilde\Gamma_{\rm int}[{\cal X},{\cal G}]$ is invariant under the same transformation. It follows that, for any diagrammatic truncation of the 2PI effective action
\beq
\tilde\Gamma_{\mu-i\alpha}[e^{i\alpha\tau Q}{\cal X},e^{i\alpha\tau Q}{\cal G}e^{-i\alpha\tau' Q}]=\tilde\Gamma_\mu[{\cal X},{\cal G}]\,.
\eeq
Assuming that there is a unique $\bar {\cal G}^{\cal X}_\mu$ for each ${\cal X}$ that solves $0=\delta\tilde\Gamma/\delta {\cal G}|_{{\cal X},\bar {\cal G}^{\cal X}_\mu}$, this implies that
\beq
\bar {\cal G}^{e^{i\alpha \tau Q}{\cal X}}_{\mu-i\alpha}=e^{i\alpha\tau Q}\bar{\cal G}^{\cal X}_\mu e^{-i\alpha\tau' Q}\,,
\eeq
and thus that the corresponding approximation $\tilde\Gamma_\mu[{\cal X}]=\tilde\Gamma_\mu[{\cal X},\bar {\cal G}^{\cal X}_\mu]$ to the 1PI effective action obeys the property
\beq
\tilde\Gamma_{\mu-i\alpha}[e^{i\alpha \tau Q}{\cal X}]=\tilde\Gamma_\mu[{\cal X}]
\eeq
from which the Silver Blaze property follows, as explained in the previous section. Thus any diagrammatic truncation of the 2PI effective action is compatible with the Silver Blaze phenomenon. We note that we have not paid much attention to the UV regularization but, clearly, any regularization that does not imply a discretisation of the time interval $[0,\beta]$ maintains the Silver Blaze property.\footnote{Note also that it is important that $\alpha$ be chosen equal to a Matsubara frequency because, strictly speaking, in equilibrium, the fields that enter as arguments of the 2PI effective action should obey periodic boundary conditions.} This is one of the reasons why, in this work, we consider a regularization that does not cut off the Matsubara sums. Nevertheless, in practical, numerical calculations, we consider a finite, however large, number of frequencies $2N_\tau+1$. In the limit $T\to 0$, we need to make sure that the largest Matsubara frequency $\omega_{N_\tau}\equiv 2\pi N_\tau T$ goes to infinity, in order for the Silver Blaze property to hold, see Fig.~\ref{fig:sb}. Indeed, if the frequency is cut-off in the zero-temperature integrals, the simple transformation property of the Lagrangian density under (\ref{eq:gauge}), which lies at the root of the Silver Blaze property, is not exactly fulfilled.

\section{Renormalization}\label{sec:renormalization}

In App.~\ref{app:renormalization}, using  the Silver Blaze property together with an expansion of the perturbative propagator around the corresponding propagator at $\mu=0$, we show that the elimination of UV divergences in perturbation theory at finite $T$ and finite $\mu$ requires the same counterterms as those needed at finite $T$ and $\mu=0$. Because the renormalization of $\Phi$-derivable approximations is just a resummed version of perturbative renormalization, we expect the two-loop $\Phi$-derivable approximation at finite $T$ and finite $\mu$ to be renormalizable using the same procedure as the one detailed in \cite{Marko:2013lxa} for the two-loop $\Phi$-derivable approximation at finite $T$ and $\mu=0$. 

\subsection{Renormalization and consistency conditions}
In the two-loop $\Phi$-derivable approximation at $\mu=0$, two bare masses $m_0$ and $m_2$ were needed because, as we mentioned above, the gap and curvature masses do not agree within a given truncation. In order to fix the two bare masses in terms of a unique renormalized mass $m_\star$ one considers the usual renormalization condition
\beq
\hat M^2_{\phi=0,T=T_\star,\mu=0}=m^2_\star
\eeq 
supplemented by a consistency condition
\beq
\hat M^2_{\phi=0,T=T_\star,\mu=0}=\bar M^2_{\phi=0,T=T_\star,\mu=0}
\eeq
which enforces the equality between the curvature and gap masses at the renormalization point. For convenience, we impose the renormalization and consistency conditions at a fixed temperature $T=T_\star$, which plays the role of the renormalization scale.

Similarly, the possibility to obtain the four-point function in three different ways \cite{Berges:2005hc,Marko:2013lxa}, which do not agree in a given $\Phi$-derivable approximation and which not all obey the crossing symmetry, requires the introduction of five bare couplings $\lambda_0^{(A,B)}$, $\lambda_2^{(A,B)}$ and $\lambda_4$. In order to fix these bare couplings in terms of a single renormalized coupling $\lambda_\star$ one considers the usual renormalization condition, supplemented by four consistency consistency conditions which impose that the three different definitions of the four-point function agree with each other, and obey crossing symmetry, at the renormalization point, that is at $T=T_\star,$ $\phi=0$ and $\mu=0$.

The explicit expressions for the bare parameters that follow from the renormalization and consistency conditions can be found in \cite{Marko:2013lxa} and are used in the present work as well. We note however that, because we have not taken $Z_0$ equal to $1$ yet, see below, one should replace in all these expressions $G_\star=1/(Q^2+m^2_\star)$ by $1/(Z_0 Q^2+m^2_\star)$.

\subsection{Field renormalization at finite $\mu$}
So far we did not discuss the need for field renormalization. In a general approximation, one needs to express the bare field in terms of the renormalized field $\varphi\to \sqrt{Z}\varphi$ and the renormalization factor $Z$ is used to absorb certain divergences. In the 2PI framework, the same reason that leads to the introduction of two different bare masses $m_0$ and $m_2$ leads to the introduction of two different renormalization factors $Z_0$ and $Z_2$, corresponding respectively to rescalings of the propagator $G$ and the field $\phi$ that enter as arguments of the 2PI effective action $\Gamma[\phi,G]$, see the beginning of Sec.~\ref{sec:2loop}.

The reason why these renormalizations are not needed in the two-loop $\Phi$-derivable approximation at $\mu=0$ is two-fold. First of all, in this approximation, the gap equation and the corresponding two-point function $\bar G_\phi$ do not involve diagrams that require the field renormalization $Z_0$. Second, even though the two-point function $\delta^2\Gamma/\delta\phi_a\delta\phi_b$ that one can construct from the corresponding approximation to the effective action $\Gamma[\phi]=\Gamma[\phi,\bar G_\phi]$ involves diagrams that do require the field renormalization $Z_2$, the analysis of \cite{Marko:2013lxa} focus on the effective potential $\gamma(\phi)$ which gives only access to the zero frequency/momentum value of this two-point function for which field renormalization is not needed.

At finite $\mu$, the situation is slightly different. It is still true that the two-point function that one obtains from solving the gap equation does not require the field renormalization $Z_0$ (because the topologies that enter the gap equation are the same as for $\mu=0$). The latter could then be fixed to $1$, but we shall keep it arbitrary (but finite) for the moment. In contrast, the two-point function that one obtains from $\gamma(\phi)$ does require the field renormalization $Z_2$, even though it still corresponds to the two-point function at zero frequency/momentum. The simplest way to understand why this is so is to consider the symmetric phase at zero temperature. There, the Silver Blaze property relates vertex functions at finite $\mu$ and vanishing external frequencies/momenta to the same vertex functions at $\mu=0$ but with non-vanishing and $\mu$-dependent external frequencies. Thus, it is no question that field renormalization is needed at finite $\mu$ to renormalize the effective potential.

We fix the renormalization factor $Z_2$ with a renormalization condition imposed on $\hat M^2_{\phi=0}$ which mimics the usual way of fixing the wave function renormalization. The renormalization factor $Z_0$ is fixed through a consistency condition which, in the exact theory limit, would ensure that the two renormalization factors converge to the same expression. The conditions we use are
\beq
\frac{d}{d\mu^2}\hat M^2_{\phi=0}\Bigg|_{T_\star,\mu=0} & = & 1-\alpha\,,\\
\frac{d}{d\mu^2}\hat M^2_{\phi=0}\Bigg|_{T_\star,\mu=0} & = & \frac{d}{d\mu^2}\bar M^2_{\phi=0}\Bigg|_{T_\star,\mu=0}\,.
\eeq
Expressing $Z_0$ and $Z_2$ from the conditions yield
\beq
Z_2 &\!\! =\!\! & \alpha\!+\!\frac{\lambda_{2}^{(A+B)}}{6}\frac{d{\cal T}[\bar {\cal D}]}{d\mu^2}\Bigg|_{T_\star,\mu=0}\!\!\!-\frac{\lambda_\star^2}{18}\frac{d{\cal S}[\bar {\cal D},
\bar {\cal D}^*,\bar {\cal D}]}{d\mu^2}\Bigg|_{T_\star,\mu=0}\,,\nonumber\\ \label{eq:Z2First}\\
Z_0 & = & \alpha+\frac{\lambda_{0}^{(A+B)}}{6}\frac{d{\cal T}[\bar {\cal D}]}{d\mu^2}\Bigg|_{T_\star,\mu=0}\,.\label{eq:Z0First}
\eeq
Carrying out the differentiations in \eqref{eq:Z2First} and \eqref{eq:Z0First}, bearing in mind that chemical potential dependence is either explicit or through the gap mass $\bar M^2_{\phi=0}$ and using the expression for $\lambda_0^{(A+B)}$ and $\lambda_2^{(A+B)}$ given in \cite{Marko:2013lxa} but including the factor $Z_0$ as explained above, one obtains
\beq
Z_2 & = & Z_0+\frac{\lambda_\star^2}{6}{\cal B}_\star[G_\star](0)\left(\frac{\partial {\cal T}[\bar {\cal D}]}{\partial \mu^2}\right)\Bigg|_{T_\star,\mu=0}\nonumber\\
& & -\,\frac{\lambda_\star^2}{18}\left(\frac{\partial {\cal S}[\bar {\cal D},\bar {\cal D}^*,\bar {\cal D}]}{\partial \mu^2}\right)\Bigg|_{T_\star,\mu=0}\label{eq:Z2withZ0}\,,\\
Z_0 & = & \alpha+\frac{\lambda_\star}{3}\frac{\partial{\cal T}[\bar {\cal D}]}{\partial\mu^2}\Bigg|_{T_\star,\mu=0}\,.
\eeq
Owing to the fact that the divergent part of ${\cal T}[{\cal D}]$ does not depend on $\mu$, see App.~\ref{app:sum}, this last expression for $Z_0$ makes it explicit that $Z_0$ is finite, as already argued. Moreover, it is easily checked that upon the rescaling $\alpha\to c\alpha$, $m^2_\star\to cm^2_\star$ and $\lambda_\star\to c^2\lambda_\star$, the renormalization factors and the squared bare masses are scaled by $c$ and the bare couplings by $c^2$. It is easily checked then that,\footnote{This is pretty clear for the explicit terms of \eqref{eq:2PI_LTA}, up to a $T$ and $\mu$-independent divergence. For the diagrammatic contributions to $\gamma_{\rm int}[\phi,G]$, we use that $E+2I=4V$ where $E$ is the number of occurences of $\phi$ in the diagram, $I$ the number of occurrences of $G$ and $V$ the number of vertices. We have then $\Gamma_{\rm int}[\phi/\sqrt{c},G/c;c^2\lambda]=c^{2V-E/2-I}\Gamma_{\rm int}[\phi,G;\lambda]=\Gamma_{\rm int}[\phi,G;\lambda]$.} up to a $T$ and $\mu$-independent divergence
\beq
\gamma[\phi/\sqrt{c},G/c;cm^2_\star,c^2\lambda_\star,c\alpha]=\gamma[\phi,G;m^2_\star,\lambda_\star,\alpha]\,,
\eeq
where we have made the dependence on the renormalized parameters and on $\alpha$ explicit. Assuming the unicity of $\bar G_\phi$ for each $\phi$ it follows that $c\,\bar G_{\phi/\sqrt{c};cm^2_\star,c^2\lambda_\star,c\alpha}=\bar G_{\phi;m^2_\star,\lambda_\star,\alpha}$ and then that
\beq
\gamma[\phi/\sqrt{c};cm^2_\star,c^2\lambda_\star,c\alpha]=\gamma[\phi;m^2_\star,\lambda_\star,\alpha]\,.
\eeq
This last relation expresses the fact that the model has only two free parameters and that $\alpha$ is arbitrary. In other words, two systems characterized by the parameters $(\alpha,m^2_\star,\lambda_\star)$ and $(c\alpha,cm^2_\star,c^2\lambda_\star)$ lead to the same physical predictions. For instance the differential pressure of the system, obtained at the minimum of $\gamma(\phi)$ is clearly the same for these two systems. Similarly, even though the curvature at the origin of the potential is scaled by $c$ when one moves from system $(\alpha,m^2_\star,\lambda_\star)$ to system $(c\alpha,cm^2_\star,c^2\lambda_\star)$, the transition line, that is the line in the $(\mu,T)$-plane where the curvature at the origin vanishes is the same for the two systems.

Because the choice of $\alpha$ is arbitrary, in what follows, we choose it such that $Z_0=1$. Hence, from now on we will omit $Z_0$. Carrying out the differentiations and limits appearing in \eqref{eq:Z2withZ0} yields
\begin{widetext}
\beq
Z_2 & = & 1-\frac{\lambda_\star^2}{18}\left[3\int_Q^{T_\star}\big(G_\star^2(Q)-4\omega_n^2G_\star^3(Q)\big)\big[{\cal B}_\star[G_\star](Q)-{\cal B}_\star[G_\star](0)\big]+4\int_Q^{T_\star}\int_K^{T_\star}\omega_n\omega_mG_\star^2(Q)G_\star^2(K)G_\star(Q+K)\right].\nonumber\\
\label{eq:Z2Explicit}
\eeq
\end{widetext}

Now that we have explained how to fix all the renormalization factors and bare parameters, we could give a detailed proof of how these parameters indeed absorb all the divergences which appear in the two-loop $\Phi$-derivable approximation at finite $T$ and finite $\mu$. Even though this is possible by extending the ideas described in App.~\ref{app:renormalization}, we shall not do this here. We will limit ourselves to provide a proof for the finiteness of the density in the broken phase, see App.~\ref{app:renormalization}, because this is a quantity we shall be dealing with later on and because its renormalization involves the renormalization factor $Z_2$, which is a new element as compared to our earlier discussion in \cite{Marko:2013lxa}.

\section{Results}\label{sec:results}

The numerical solution of the field and gap equations is obtained iteratively, as mentioned earlier. We extend the approach used in \cite{Marko:2012wc} which exploits Fourier analysis and rotation invariance. We mention however that the skew component $\bar G_A$ is odd under the transformation $\omega_n\to -\omega_n$ which leads to certain complications in the practical implementation of our approach. More precisely, even though it is straightforward to implement the numerical convolution of an even function with an odd function, this requires a sampling of the odd function different from the one used for the even function. This leads to a certain loss information when evaluating the propagators using (\ref{eq:GT}) and (\ref{eq:GA}), because $\Delta$ involves both even and odd functions which are not sampled in the same way. To avoid a loss of information, we choose to rewrite our equations in terms of a function $\bar g_A$ such that $\bar G_A=\omega_n \bar g_A$. This function $\bar g_A$ is even under the transformation $\omega_n\to -\omega_n$. The only convolutions that we have to consider are thus convolutions involving two even functions, for which we can use the routines described in \cite{Marko:2012wc}. Our results do not depend on the particular value chosen for $\bar g_A(\omega_n=0)$. We also note that \eqref{eq:Z2Explicit} contains frequency-odd sum-integrands, which we treat using $\omega_n\omega_m=[(\omega_n+\omega_m)^2-\omega_n^2-\omega_m^2]/2.$

\subsection{Transition line}
As already mentioned, the transition line $T_{c}(\mu)$ is obtained from the condition:
\beq
\hat M^2_{\phi=0;T=T_{c}(\mu),\mu}=\mu^2\,,
\label{eq:defTc}
\eeq
where the curvature mass is given in \eqref{eq:symm_curvature} in terms of the gap mass $\bar M^2_{\phi=0}$. We mention that the curvature at the origin of the potential is defined only if the gap equation \eqref{eq:M20} for $\bar M^2_{\phi=0}$, which enters the expression for $\bar M^2_{\phi=0}$, admits a solution. This gap equation can be written as $0=f(\bar M^2_{\phi=0})$ with $f(M^2)=-M^2+m^2_0+(\lambda_0^{(A+B)}/6){\cal T}[{\cal D}]$ where $f(M^2)$ is defined for $M^2\geq\mu^2$. We thus need to study the zeroes of $f(M^2)$ for $M^2\geq\mu^2$. A direct calculation shows that
\beq\label{eq:der}
f'(M^2)=-1-\frac{\lambda_0^{(A+B)}}{6}{\cal B}[{\cal D}]<0\,,
\eeq
where it is implicitly assumed that we keep our cut-off below the Landau scale,\footnote{The presence of the Landau pole makes renormalization meaningful only if there is a large separation between the physical scales and the Landau scale. For this reason, we restrict our analysis to parameter values such that the Landau scale is much larger than our renormalization scale $T_\star$.} that is we have always both $\lambda_0^{(A)}>0$ and $\lambda_0^{(B)}>0$, see our discussion in \cite{Marko:2012wc,Marko:2013lxa}. From (\ref{eq:der}), it follows that $f(M^2)$ decreases from $f(\mu^2)$ to $-\infty$ and thus there is a solution if and only if $f(\mu^2)\geq 0$. The condition $f(\mu^2)=0$, or equivalently $\bar M^2_{\phi=0}=\mu^2$ defines a line $\bar T_c(\mu)$ below which the gap equation at $\phi=0$ has no solution. It would correspond to the critical line $T_c(\mu)$ if the gap and curvature masses were to coincide. If both $\bar T_c(\mu)$ and $T_c(\mu)$ exist then by definition $T_c(\mu)\geq\bar T_c(\mu)$ at any $\mu$, as the gap equation at vanishing field loses its meaning for $T<\bar T_c(\mu)$.

\begin{center}
\begin{figure}
\includegraphics[width=0.45\textwidth]{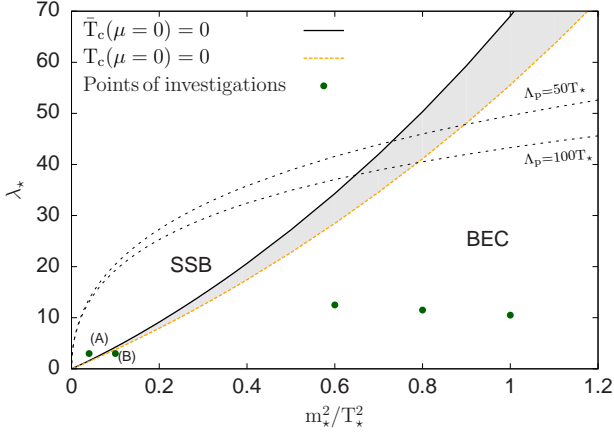}\caption{The parameter space divided into SSB and BEC regions according to the existence of the critical temperatures $T_c(\mu=0)$ and $\bar T_c(\mu=0)$. Systems for which both temperatures exist show SSB whereas systems for which neither exist show BEC. The greyed region is an artefact of the two-loop 2PI approximation and has no physical interpretation. The labeled dashed lines denote the limits over which the Landau pole is smaller than $50T_\star$ and $100 T_\star$ respectively. The filled circles denote points where we carried out further investigations. At points (A) and (B) the iso-density lines are compared to the corresponding results obtained in the Hartree-Fock approximation (see Sec.~\ref{sec:isorho}), while the unmarked points are used in the lattice comparison (see Sec.~\ref{sec:lattice}).\label{fig:plane}}
\end{figure}
\end{center}

There are three different cases based on the value of the two critical temperatures at $\mu=0$. We call the spontaneous symmetry breaking (SSB) case, the one where both $T_c$ and $\bar T_c$ are defined and larger than zero for $\mu=0$. We call the Bose-Einstein condensation (BEC) case the one where neither $T_c$ nor $\bar T_c$ exists at vanishing chemical potential, but they appear at critical values $\mu_c$ and $\bar \mu_c$ respectively. There is also a third case when $T_c>0$ exists at $\mu=0$ however $\bar T_c(\mu=0)$ is not defined. Since this is just an artefact of the two-loop $\Phi$-derivable approximation, we will not discuss this case any further. The different cases divide the $m_\star^2-\lambda_\star$ parameter plane as shown in Fig.~\ref{fig:plane}. The typical behaviour of the $T_c(\mu)$ and $\bar T_c(\mu)$ curves in both cases is shown in Fig.~\ref{fig:isorho}.

\subsection{Iso-density lines}\label{sec:isorho}

The density is defined as
\beq
\rho=\frac{1}{VZ}{\rm Tr}\,Q\,e^{-\beta(H-\mu Q)}=\frac{1}{\beta V}\frac{\partial\ln Z}{\partial\mu}\,.
\eeq
Since the path integral formula for $\ln Z$ depends on $\mu$ only through the quadratic part of the Lagrangian density we obtain
\beq
\rho & = & \mu\langle\varphi^2_1(0)+\varphi^2_2(0)\rangle+i\langle\dot\varphi_2(0)\varphi_1(0)\rangle-\langle\dot\varphi_1(0)\varphi_2(0)\rangle\nonumber\\
& = & \mu \bar\phi^2+\mu\int_Q (\bar G_L(Q)+\bar G_T(Q))-2\int_Q \omega_n \bar G_A(Q)\,,\nonumber\\
\eeq
where we have used (\ref{eq:gg}). In the symmetric phase, we have $\bar\phi=0$ and $\bar G_L=\bar G_T=\bar G_I$ and it is more convenient to express the density in terms of $\bar {\cal D}=\bar G_I-i\bar G_A$ and $\bar {\cal D}=\bar G_I+i\bar G_A$. Writing $2\bar G_I=\bar {\cal D}+\bar {\cal D}^*$ and $2\bar G_A=i(\bar {\cal D}-\bar {\cal D}^*)$, we arrive at
\beq
\rho=\int_Q\left[\frac{i\omega_n+\mu}{-(i\omega_n+\mu)^2+\varepsilon_q^2}-\frac{i\omega_n-\mu}{-(i\omega_n-\mu)^2+\varepsilon_q^2}\right]
\eeq
which is easily computed to be
\beq
\rho=\int_q\left[\frac{1}{e^{\beta(\varepsilon_q-\mu)}-1}-\frac{1}{e^{\beta(\varepsilon_q+\mu)}-1}\right],
\eeq
with $\varepsilon_q=\sqrt{q^2+\bar M^2_{\phi=0}}$. The density in the symmetric phase is thus finite provided $\bar M^2_{\phi=0}$ is. That this is the case follows immediately from the fact that the divergent part of the tadpole integral in the rhs of (\ref{eq:M20}) does not depend on $\mu$, see App.~\ref{app:sum}. The renormalization in the broken phase, using the renormalization factor $Z_2$ is done in App.~\ref{app:renormalization}.
\vglue2mm

\begin{center}
\begin{figure}
\includegraphics[width=0.45\textwidth]{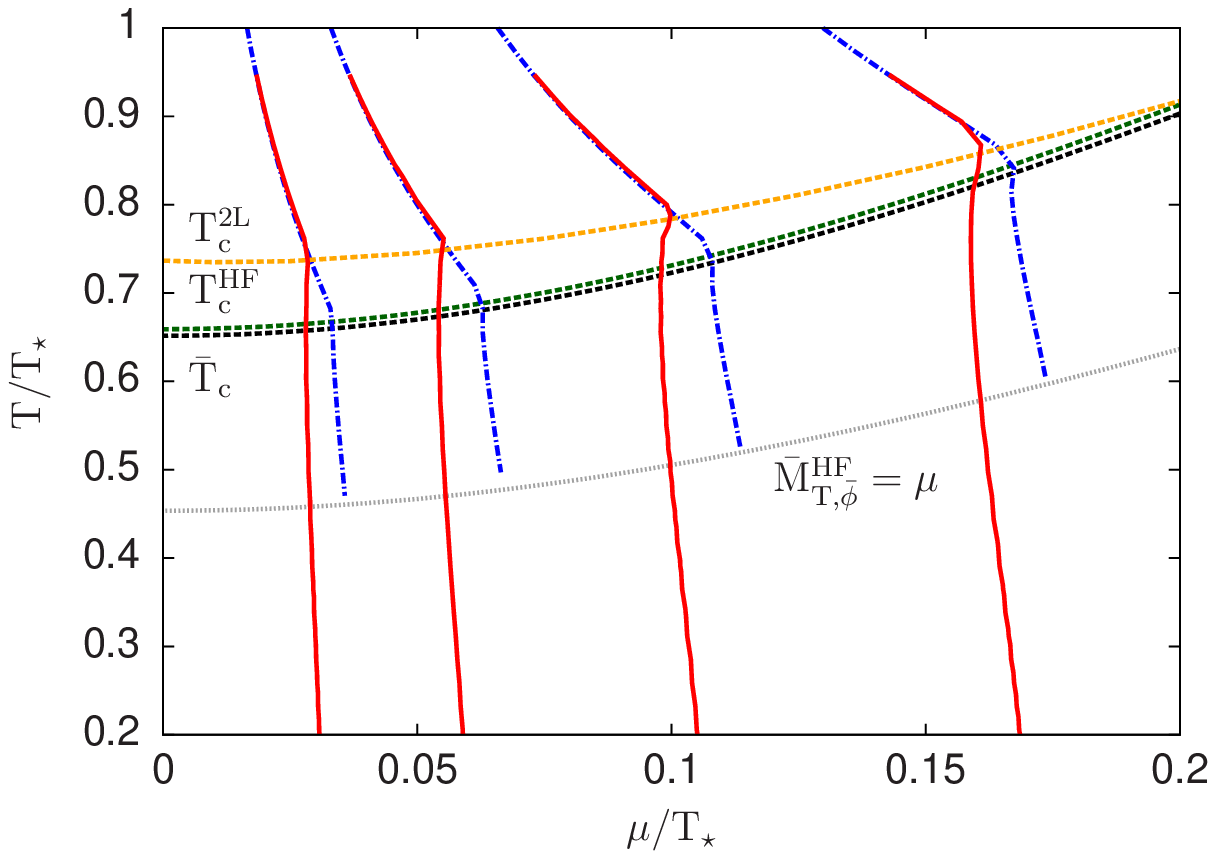}\vspace{0.02\textwidth}\\\includegraphics[width=0.45\textwidth]{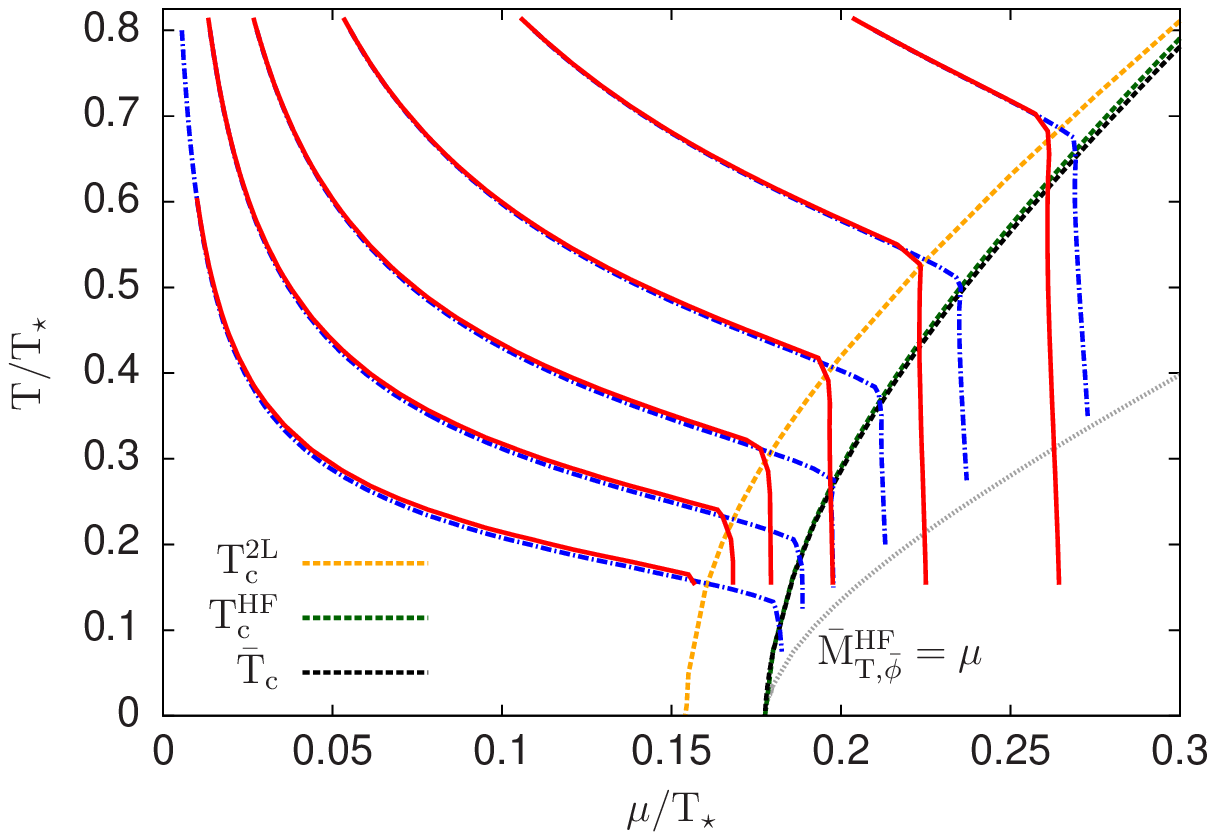}
\caption{Phase diagrams and iso-density lines in the SSB (top panel) and BEC (bottom panel) cases. The parameter values chosen correspond to points (A) and (B) of Fig.~\ref{fig:plane} respectively, that is $(m_\star^2/T_\star^2,\lambda_\star)=(0.04,3)$ for (A) and $(m_\star^2/T_\star^2,\lambda_\star)=(0.1,3)$ for (B). The full lines are the isodensity lines in the two-loop approximation, the dot-dashed lines are their counterpart in the Hartree-Fock approximation. The dashed lines are the critical temperature curves, while the dotted lines are the limiting lines in the Hartree-Fock approximation, under which the gap equation has no solution at the would-be minimum of the potential. The chosen $\rho$ values from left to right are \{$\rho/T_\star^3=0.05\,,0.01\,,0.02\,,0.04$\} in the SSB case, and \{$0.001\,,0.0025\,,0.05\,,0.01\,,0.02\,,0.04$\} in the BEC case.\label{fig:isorho}}
\end{figure}
\end{center}

We took two example points (shown in Fig.~\ref{fig:plane}) in the $m_\star^2/T_\star^2-\lambda_\star$ plane, one in the SSB region $(m_\star^2/T_\star^2=0.04\,,\lambda_\star=3)$ and one in the BEC region $(m_\star^2/T_\star^2=0.1\,,\lambda_\star=3)$. We located the $T_c(\mu)$ and $\bar T_c(\mu)$ curves, then determined density values on a grid from which we interpolated iso-density lines, i.e. lines of constant density. The results are compared to the same quantities obtained in the lower Hartree-Fock approximation in Fig.~\ref{fig:isorho}. Deep in the symmetric phase the Hartree-Fock and the two-loop approximations are equivalent for the density. The small difference is caused by the error of the interpolation. The main difference is in the ``breakpoint'' of the curves, which is determined by the location of the phase transition, which is different in the two approximations. Furthermore in the Hartree-Fock approximation the density lines are in fact discontinuous as the transition is first order, however the weakness of the phase transition makes this negligible for the purpose of this comparison. In the broken phase lines of constant $\rho$ are almost lines of constant $\mu$ in both approximations, although the values are different.

\begin{center}
\begin{figure}
\includegraphics[width=0.45\textwidth]{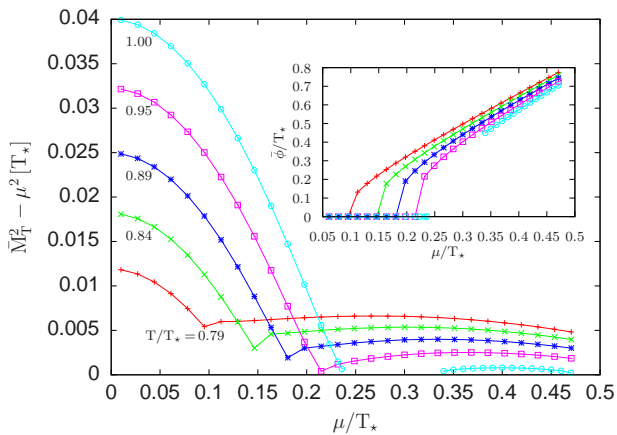}\caption{The difference $\bar M_T^2-\mu^2$ as a function of $\mu$ for several
temperatures. The integrals in the coupled field and gap equations lose
their meaning for $\bar M_T^2-\mu^2<0$. This happens for the largest
temperature in the region where the solution is missing, while for the
smaller temperatures arbitrary many points could be taken in the shown
region, jumps are only consequences of the chosen resolution. The inset
shows the corresponding $\bar\phi(\mu)$ curves for each
$\bar M_T^2(\mu)-\mu^2$ curve.\label{fig:GSprob}}
\end{figure}
\end{center}

We must note that there are certain regions of the $\mu-T$ plane, which cannot be accessed by our current approximation. For large temperature and chemical potential in the broken phase the coupled field and gap equations lose their solution. There seem to be two connected yet distinguishable reasons behind the loss of solution. The first one can be summarized as follows. For $\mu-T$ pairs under the $\bar T_c(\mu)$ curve, by definition of $\bar T_c$, there exists a $\phi_c(\mu,T)$ such that for $\phi<\phi_c$ the gap equations lose their solution, as the difference $\bar M_T^2-\mu^2$ would become negative, rendering integrals containing $\bar G_T$ meaningless. The solution of the coupled field and gap equations is lost if $\phi_c$ reaches $\bar\phi$ at a certain $\mu-T$ point. The same mechanism prevents us from solving the Hartree-Fock approximation in the regions bordered by the dotted grey lines of Fig.~\ref{fig:isorho}. The other reason is connected with the infrared sensitivity of diagrams included in our approximation. As $\bar\phi$ approaches $\phi_c$, $\bar M_T^2-\mu^2$ gets closer and closer to zero. However the zero external momentum value of the bubble diagram with two transverse propagators, appearing in the longitudinal gap equation \eqref{eq:ML}, diverges in this limit. This leads to a loss of solution, however in a slightly different way as in the first case. As a consequence the Goldstone theorem may not be fulfilled in any way in the two-loop approximation, neither in any other approximations where these infrared divergences are not tamed by further resummations. Fig.~\ref{fig:GSprob} shows the effective transverse gap mass as a function of the chemical potential for several high temperatures, together with $\bar\phi$ in the inset, at point (A) of the parameter space, to illustrate the loss of solution.

\subsection{Comparison to lattice results}\label{sec:lattice}
The $O(2)$ model at finite density has been studied on the lattice in \cite{Gattringer:2012df} using an appropriate lattice discretisation similar to (\ref{eq:LE3}) and the flux tube representation of the partition function which solves the sign problem in this case. These results have been quite remarkably reproduced using the extended mean-field method \cite{Akerlund:2014mea}, a semi-analytical approach that goes beyond mean field. This latter approach uses the same lattice discretisation as in \cite{Gattringer:2012df} with of course the same bare parameters in lattice units. 

It is for us difficult to compare to these results because we did not use the lattice action as our starting point.\footnote{Unfortunately, we were not aware of reference \cite{Gattringer:2012df} when starting this work.} A comparison would only make sense if the cut off (inverse lattice spacing) used in \cite{Gattringer:2012df} or \cite{Akerlund:2014mea} was large with respect to the other scales in the problem, in which case renormalizability ensures that the physics at small momentum scales should not depend much on the starting microscopic theory within the same class of universality. However \cite{Gattringer:2012df} or \cite{Akerlund:2014mea} consider lattice spacings $a$ comparable to physical scales (for instance the density is plotted as a function of $\mu$ in two cases corresponding respectively to $a\mu_c\approx 1.15$ and $a\mu_c\approx 0.18$). We could of course lower our cutoff but we would then be sensitive to the microscopic details between (\ref{eq:LE2}) and the lattice discretised version of (\ref{eq:LE3}).

\begin{center}
\begin{figure}
\includegraphics[width=0.45\textwidth]{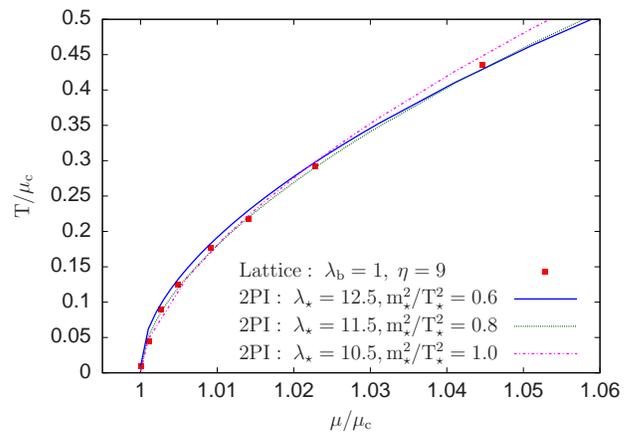}
\caption{To fix the 2PI parameters $m_\star^2$ and $\lambda_\star$ for a comparison to lattice results, we
choose to reproduce the $T_c(\mu)$ curve obtained with lattice simulations in \cite{Gattringer:2012df}
parametrized by the bare quantities $\lambda_b=1$ and $\eta=9$ (the shown data was read off approximately
from Fig.~5 of \cite{Gattringer:2012df}). This however does not fix our parameters completely, there is still
some arbitrariness left. Both axes are scaled by the critical chemical potential value at zero temperature,
$\mu_c$, to compare dimensionless quantities.\label{fig:Tc_curve_vs_Latt}}
\end{figure}
\end{center}

Even though a quantitative comparison does not really make sense, we tried the following qualitative comparison. We chose the parameters of the two-loop $\Phi$-derivable approximation, such that the $T_c(\mu/\mu_c)/\mu_c$ curve shown on Fig.~5 of \cite{Gattringer:2012df} with the bare parameters $\lambda_b=1$ and $\eta=9$ in lattice units, is reproduced. We show the comparison on Fig.~\ref{fig:Tc_curve_vs_Latt}. We chose the parameters such, that the curve is reproduced by our $T_c(\mu)$ curve, while we could also use $\bar T_c(\mu)$ for this purpose. However, as the minimum of the potential changes at $T_c$ and not at $\bar T_c$ we choose the former. Note that this procedure does not fix our parameters completely, as we can only reproduce the phase transition curve up to some accuracy to which corresponds some patch in the parameter space. Instead of really finding the boundaries for a certain error we choose three distinctly different parameter sets ($[m_\star^2/T_\star^2,\lambda_\star]:\,[0.6,12.5],\,[0.8,11.5],\,[1.0,10.5]$) to estimate the dependence of the results on the remaining arbitrariness.

\begin{center}
\begin{figure}
\hspace*{0.45cm}
\includegraphics[width=0.35\textwidth, angle=-90]{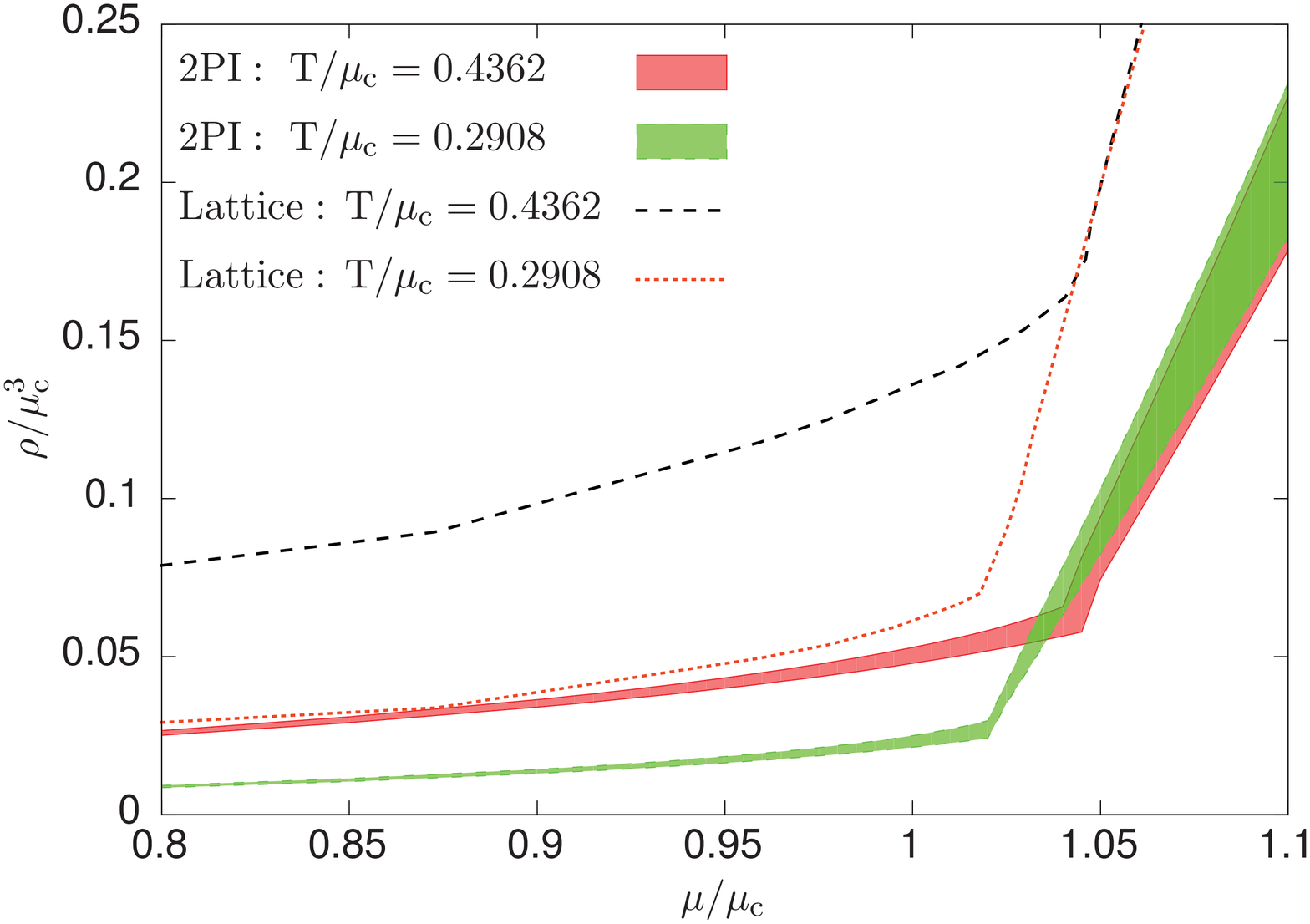}
\caption{The dimensionless density $\rho/\mu_c^3$ as a function of $\mu/\mu_c$ for two different temperatures. The shaded regions show the two-loop 2PI results including the uncertainty our parameter fixing carries, while the lattice data was approximately read off from Fig.~6 of \cite{Gattringer:2012df}.\label{fig:density_vs_Latt}}
\end{figure}
\end{center}

At the chosen parameters we compare the dimensionless quantity $\rho/\mu_c^3$. The results are shown in Fig.~\ref{fig:density_vs_Latt}. We find that our density is smaller than the lattice values. Again, we do not expect quantitative agreement but we note that, interestingly, the difference between our results and the lattice results is a constant scaling factor which does not depend on the chemical potential and depends mildly on the temperature. Furthermore the dependence of the density on the choice of parameters is almost negligible in the symmetric phase, while somewhat stronger in the broken phase, but the apparent slope of the curves still does not come close to the lattice results.

\section{Conclusions}
We studied the charged scalar $O(2)$ model with quartic interaction at finite temperature and
non-zero chemical potential within the two-loop $\Phi$-derivable approximation. The fact that the Euclidean action is complex in the presence of a real chemical potential poses certain problems in the 2PI formalism (and more generally in any approach based on Legendre transforms) which are reminiscent of the sign problem on the lattice. We discussed these issues and solved some of them in the particular equilibrium context of this work.

We solved the approximation by extending the numerical approach 
of \cite{Marko:2012wc} which exploits rotation invariance and the fast Fourier transform algorithm. Depending on the values of the parameters $m_\star^2$ and $\lambda_\star$, the system displays
either spontaneous symmetry breaking or Bose-Einstein condensation. From the chemical
potential and temperature dependence of the effective potential we conclude that both phase
transitions are of the second order type. In the Bose-Einstein condensation case our numerical results are consistent with
the Silver Blaze property. This comes as no surprise because the approximation, the regularization and the discretization that we use respect a particular transformation rule of the Euclidean action under certain gauge transformations of the field, which we showed to be at the root of the Silver Blaze property and its generalization to higher vertex functions. This generalization allowed us to argue in particular that the model at finite chemical potential can be renormalized using the same counterterms as those needed at zero chemical potential.

We compared with the lattice results of \cite{Gattringer:2012df} even though a quantitative comparison is not possible since cutoff effects are not small at the lattice spacings used in \cite{Gattringer:2012df} and also because in our study we are not using the lattice action. Nevertheless we
can choose parameters to reproduce the phase transition line on the $\mu-T$ plane to good
accuracy. At the parameters fulfilling this parametrization criterion, we compared the density as a function of the chemical potential at several temperatures, where we reproduce the qualitative features of the curves.

\begin{acknowledgments}
Zs.~Sz\'ep was supported by the Hungarian Research Fund (OTKA) under Contract No. K104292.
\end{acknowledgments}

\appendix

\section{Complex actions}\label{app:complex}
In the case of a real-valued action with a real-valued (multi-component) field $\varphi$ coupled to a real-valued source $J$, the Legendre transformation, which allows to obtain the 1PI effective action $\Gamma[\phi]$ from the generating functional $W[J]$, maps the real-valued source $J$ into the real-valued Legendre variable $\phi$. If the action becomes complex, as it is the case in the presence of a finite chemical potential, even though it is still natural to consider a real-valued source because the field over which one integrates remains real, the Legendre transformation maps this source into a complex-valued Legendre variable. Moreover, the components of this variable are constrained since they should correspond, by inverse Legendre transformation, to a source with real-valued components.

Let us illustrate these points by computing the 1PI effective action of the theory (\ref{eq:LE1}) in the limit of zero coupling  (in this case the bare mass $m^2_{\rm b}$ is finite and we write it $m^2$ in what follows). We first determine the generating functional $W_\mu[J]$ in the presence of a real-valued source $J$. This boils down to the evaluation of the Gaussian integral
\beq
e^{W_\mu[J]}=\int {\cal D}\varphi\,e^{-\frac{1}{2}\varphi^{\rm t}G^{-1}_0\varphi+J^{\rm t}\varphi}\,,
\eeq
where we have used a schematic notation in terms of infinite vectors and matrices whose coordinates are labelled not only by the internal indices of the field but also by space and time variables, and periodic boundary conditions are understood even though our notation does not make it explicit. The usual way to deal with this integral is to redefine the field as $\varphi\to \varphi+G_0J$. This gives
\beq\label{eq:step}
e^{W_\mu[J]}=e^{\frac{1}{2}J^{\rm t}G_0J}\int {\cal D}\varphi\,e^{-\frac{1}{2}\varphi^{\rm t}G^{-1}_0\varphi}\,,
\eeq
where we have used the fact that $G_{0,ab}^{-1}(x,x')=G_{0,ba}^{-1}(x',x)$, as it can be readily checked using Eq.~(\ref{eq:G0op}). We have cheated a bit in writing (\ref{eq:step}) because $G_0^{-1}$ being complex the change of variables $\varphi\to\varphi+G_0J$ changes the region over which the fields  are integrated. Note that this only affects the integral in (\ref{eq:step}) which does not depend on the sources. Moreover, it can still be argued that this integral is equal, up to some factor, to $({\rm det}\,G_0^{-1})^{-1/2}$. Now that we know the explicit dependence of $W_\mu[J]$ with respect to the sources, we can obtain the explicit relation between the source and the variable $\phi$ that enters the 1PI effective action. It is $\phi=G_0J$ or $J=G_0^{-1}\phi$, which we write more explicitly as
\beq
J_1 & = & -\left(\frac{\partial^2}{\partial\tau^2}+\Delta-m^2+\mu^2\right)\phi_1-2i\mu\frac{\partial\phi_2}{\partial\tau}\,,\label{eq:one1}\\
J_2 & = & -\left(\frac{\partial^2}{\partial\tau^2}+\Delta-m^2+\mu^2\right)\phi_2+2i\mu\frac{\partial\phi_1}{\partial\tau}\,.\label{eq:two2}
\eeq
These formulae show very clearly that, in general, $\phi_1$ and $\phi_2$ cannot be real if the sources $J_1$ and $J_2$ are taken real. Nevertheless, we can pursue the determination of the 1PI effective action which reads
\beq
\Gamma_\mu[\phi] & = & J^{\rm t}\phi-W_\mu[J]\nonumber\\
& = & \frac{1}{2}\phi^{\rm t}G_0^{-1}\phi+\frac{1}{2}{\rm Tr}\,\ln G_0^{-1}\,.
\eeq
We have thus obtained the usual formula. The only change with respect to the case where the action is real is that, as one varies the real-valued components of the source $J$, the components of the Legendre variable $\phi$ take complex values, constrained by the fact that the components of the original source are real.

One way to avoid the presence of this constraint is to consider a complex-valued source $J$ in which case $\phi$ is complex-valued with unconstrained components. However, in some situations of interest, it is not necessary to consider such an extension of the source because $\phi$ can remain real-valued despite the fact that the action is complex. For instance, according to (\ref{eq:one1}) and (\ref{eq:two2}), a situation where this is true in the free theory is that of a static system. In fact this holds also in the interacting case as we now argue. The relation between the source $J$ and the Legendre variable $\phi$ is nothing but
\beq\label{eq:phi_def}
\phi(x)\equiv \frac{\int{\cal D}\varphi\,\varphi(x)\,e^{-\int_x\left({\cal L}_E-J^{\rm t}\varphi\right)}}{\int{\cal D}\varphi\,e^{-\int_x\left({\cal L}_E-J^{\rm t}\varphi\right)}}\,.
\eeq
Taking the complex conjugate of Eq.~(\ref{eq:phi_def}) and using that the sources are chosen real, one obtains
\beq
(\phi(x))^*\equiv \frac{\int{\cal D}\varphi\,\varphi(x)\,e^{-\int_x\left({\cal L}^*_E-J^{\rm t}\varphi\right)}}{\int{\cal D}\varphi\,e^{-\int_x\left({\cal L}^*_E-J^{\rm t}\varphi\right)}}\,.
\eeq
We can now consider a change of variables that corresponds to ``time reversal'' defined here as $\varphi(x)\to\varphi(Tx)$ with $\smash{x\equiv(\tau,\vec{x})}$ and $\smash{Tx\equiv(\beta-\tau,\vec{x})}$. Upon such a change of variables, the action is changed into its complex conjugate while the source term remains the same, because we consider a static source $J$. It follows that $(\phi(x))^*=\phi(Tx)$. Now since time-translation invariance is assumed not to be broken, $\phi(x)$ does not depend on $\tau$ and is therefore real, as announced.

Consider now the case of the 2PI effective action which is obtained as the Legendre transformation of the generating functional $W_\mu[J,K]$ in the presence of local and bilocal sources. We consider a homogeneous system and thus restrict to translation invariant sources $J$ and $K$. The propagator $\bar G_\phi(x-y)$ reads
\beq
\bar G_\phi(x-y) & \equiv & \frac{\int{\cal D}\varphi\,\varphi(x)\varphi^{\rm t}(y)\,e^{-\int_x\left({\cal L}_E-J^{\rm t}\varphi\right)}}{\int{\cal D}\varphi\,e^{-\int_x\left({\cal L}_E-J^{\rm t}\varphi\right)}}- \phi(x)\phi^{\rm t}(y)\,.\nonumber\\
\eeq
A similar argument as above shows that $\bar G_\phi(x-y)^*=\bar G_\phi(y-x)$. In Fourier space this leads to $\bar G_\phi(Q)^*=\bar G_\phi(Q)$. This means that we can restrict the 2PI effective action to propagators whose Fourier transform $G(Q)$ is real.\\

Another difficulty related to the fact that the Euclidean action is complex in the presence of a finite chemical potential concerns the convexity of $W_\mu[J]$. First of all we note that, if the sources are arbitrary, the question of convexity is not well posed because $W_\mu[J]$ is not even real. In the case of homogeneous sources, where $W_\mu[J]$ is real, determining whether $W_\mu[J]$ is convex can be a difficult task. For any real vector $\eta$, one shows that $\eta_a\eta_b\partial^2W/\partial J_a\partial J_b$ is equal to
\beq
\frac{\int{\cal D}\varphi \left(\int_x\eta_a(\varphi_a(x)-\phi_a)\right)^2e^{-\int_x\left({\cal L}_E-J^{\rm t}\varphi\right)}}{\int{\cal D}\varphi\,e^{-\int_x\left({\cal L}_E-J^{\rm t}\right)}}\,,
\eeq
that is the expectation value of a positive quantity ${\cal P}\equiv\left(\int_x\eta_a(\varphi_a(x)-\phi_a)\right)^2$, with however a complex weight! We can rewrite this average in terms of a real weight
\beq\label{eq:cos}
\frac{\int{\cal D}\varphi\, {\cal P}\,e^{-\int_x\left({\cal L}^{\mu=0}_E-\mu^2\varphi^2-J^{\rm t}\varphi\right)}\cos\Big(\mu\int_0^\beta d\tau\,Q(\tau)\Big)}{\int{\cal D}\varphi\,\,e^{-\int_x\left({\cal L}^{\mu=0}_E-\mu^2\varphi^2-J^{\rm t}\varphi\right)}\cos\Big(\mu\int_0^\beta d\tau\,Q(\tau)\Big)}\nonumber\\
\eeq
where we have again used a ``time reversal'' transformation and
\beq
Q(\tau)=\int d^3x\big[\dot\varphi_2\varphi_1-\dot\varphi_1\varphi_2\big]
\eeq
is the charge associated with a given configuration $\varphi$.  If $\mu$ is very small, the only configurations which make the oscillating cosine function in (\ref{eq:cos}) deviate from $1$ are those for which the time integrated charge $\int_0^\beta d\tau\,Q(\tau)$ is large but these configurations are suppressed by the real exponential factor in the measure. For non small values of $\mu$, the situation is more subtle and the convexity of $W_\mu[J]$ could rely on cancellations between differently charged field configurations. This is what we referred to in the main text as the ``small sign problem''.

\section{Structure of $\bar G_{ab}(Q)$}\label{app:tensor}
Let us see to which extent the $SO(2)$ invariance of (\ref{eq:LE1})  constrains the form of $\bar G^\phi$. First of all, the propagator $\bar G^\phi$ is covariant upon $SO(2)$ rotations of $\phi$:
\beq\label{eq:covariance}
\bar G^{R\phi}_{ab}(Q)=R_{ac}R_{bd}\bar G^\phi_{cd}(Q)\,.
\eeq
We remark that despite the fact that $\bar G_{ab}(Q)=\bar G_{ba}(-Q)$, we do not have a priori $\bar G_{ab}(-Q)=\bar G_{ab}(Q)$ since parity is conserved but not time-reversal. Thus $\bar G_{ab}(Q)$ is not necessarily symmetric under $a\leftrightarrow b$. Let us then decompose $\bar G_{ab}(Q)$ into a symmetric part $\bar G^S_{ab}(Q)$ and an antisymmetric part $\bar G^A_{ab}(Q)$. They both obey (\ref{eq:covariance}). The antisymmetric part reads $\bar G_A(Q)\,\varepsilon_{ab}$ and, because $\varepsilon_{ab}$ is $SO(2)$ invariant, (\ref{eq:covariance}) implies that $\bar G_A(Q)$ is invariant under rotations of $\phi$, that is it depends only on $\phi^2$. On the other hand, because it is real, the symmetric part $\bar G^S_{ab}(Q)$ admits two orthogonal eigenvectors. If we write one of them as $S^\phi\phi$ with $S^\phi\in SO(2)$, we obtain
\beq
\bar G^S_{ab}(Q)=S^\phi_{ac}S^\phi_{bd}\Big(\bar G_L(Q) P^L_{cd}+\bar G_T(Q) P^T_{cd}\Big).
\eeq
Using that $SO(2)$ is abelian and (\ref{eq:covariance}) one shows that $S$ does not depend on $\phi$ and that $\bar G_L$ and $\bar G_T$ depend on $\phi$ only through $\phi^2$. We have thus shown that
\beq
\bar G_{ab}(Q)=S_{ac}S_{bd}\Big(\bar G_L(Q) P^L_{cd}+\bar G_T(Q) P^T_{cd}\Big)+\bar G_A(Q)\varepsilon_{ab}\,.\nonumber\\
\eeq
This is the most general form of the propagator compatible with (\ref{eq:covariance}). Note the presence of the arbitrary $SO(2)$ matrix $S$. The reason why the presence of this arbitrary matrix $S$ is compatible with (\ref{eq:covariance}) is that $SO(2)$ is abelian. 

In order to fix even further the form of the propagator we need to use some extra information. This is provided by the form of the 2PI effective action (\ref{eq:2PI}). 
From the 2PI effective action, we obtain the gap equation (taking $Z_0=1$)
\beq
\bar G^{-1}_{ab}(K)=(K^2-\mu^2)\delta_{ab}-2\mu\omega_n\varepsilon_{ab}+\bar M^2_{ab}(K)\,.\nonumber\\
\eeq
with
\begin{widetext}
\beq\label{eq:gap}
\bar M^2_{ab}(K) & = & m^2_0\delta_{ab}+\frac{\lambda^{(A)}_2}{12}\phi^2\delta_{ab}+\frac{\lambda^{(B)}_2}{6}\phi_a\phi_b+\frac{\lambda^{(A)}_0}{12}\delta_{ab}\int_Q^T {\rm tr}\,\bar G(Q)+\frac{\lambda^{(B)}_0}{6}\int_Q^T\bar G_{ab}(Q)\nonumber\\
& - & \frac{\lambda^2_\star}{72}\phi_a\phi_b \int_Q^T {\rm tr}\,\big[\bar G(Q)\bar
G(Q+K)\big]-\frac{\lambda^2_\star}{36} \int_Q^T \phi\,\bar G(Q)\phi\,\bar G_{ab}(Q+K)\nonumber\\
& - &  \frac{\lambda^2_\star}{36}\phi_a\int_Q^T \big[\bar G(Q)\bar
G(Q+K)\phi\big]_b-\frac{\lambda^2_\star}{36}\phi_b\int_Q^T \big[\bar G(Q+K)\bar
G(Q)\phi\big]_a\nonumber\\
& - & \frac{\lambda^2_\star}{36} \int_Q^T \big[\phi\,\bar G(Q)\big]_a\big[\phi\,\bar G(Q+K)\big]_b\,,
\eeq
\end{widetext}

It is then not difficult to check that these equations are compatible with the above decomposition only if $S=1$.

\section{Renormalization at finite $\mu$}\label{app:renormalization}

\subsection{Perturbation theory}
At zero temperature and zero chemical potential, to each diagram of the perturbative expansion, the Forest formula associates a series of counterterm diagrams that eliminate all the divergences of the original diagram. It is well known that the same diagram considered at finite temperature is renormalized by the same series of diagrams extended to finite temperature but with the same counterterms as those determined at zero temperature. Even though this feature is expected on physical grounds,\footnote{Since counterterms are mere redefinitions of the parameters of the microscopic theory, it should be possible to find schemes in which they do not depend on external parameters such as the temperature or the chemical potential.} checking it in the imaginary time formalism usually requires the explicit evaluation of the Matsubara sums for each diagram, which makes the proof cumbersome. If one adds the chemical potential, it is possible to generalize the previous result. One way is again to evaluate explicitly the Matsubara sums as in the examples discussed in App.~\ref{app:sum}. However, as we now show the situation is in fact simpler because the discussion can be carried out without evaluating the Matsubara sums explicitly.

Let us consider an example first. At zero chemical potential, the sunset diagram $(-\lambda^2/18){\cal S}[{\cal D}_{\mu=0}](K)$ is renormalized by the following series of counterterm diagrams
\beq\label{eq:finiteT}
K^2\delta Z+\delta m^2+\frac{\delta\lambda}{3} {\cal T}[{\cal D}_{\mu=0}]
\eeq
with $\delta Z$, $\delta m^2$ and $\delta\lambda$ determined at zero temperature. In particular $\delta\lambda$ is needed to absorb the divergent part of the bubble diagram $-(\lambda^2/2){\cal B}[{\cal D}_{\mu=0}]$. We would like to show that at finite $\mu$, the sunset diagram $(-\lambda^2/18){\cal S}[{\cal D},{\cal D}^*,{\cal D}](K)$, which is a contribution to the self-energy corresponding to $\langle\Phi^*(x)\Phi(y)\rangle$, is renormalized by the series of counterterm diagrams
\beq\label{eq:finitemu}
((\omega_n+i\mu)^2+k^2)\delta Z+\delta m^2+\frac{\delta\lambda}{3} {\cal T}[{\cal D}]
\eeq
with the same counterterms as in (\ref{eq:finiteT}). The reason why $\delta Z$ is multiplied by $(\omega_n+i\mu)^2+k^2$ and not just $K^2$ is that the chemical potential modifies the quadratic part of the Euclidean action. 

Because we already know that (\ref{eq:finiteT}) renormalizes the sunset diagram at $\mu=0$, it is enough to show the difference $(-\lambda^2/18)[{\cal S}[{\cal D},{\cal D}^*,{\cal D}](K)-{\cal S}[{\cal D}_{\mu=0}](K)]$ is renormalized by the difference between the counterterm diagrams of (\ref{eq:finitemu}) and those of (\ref{eq:finiteT}), that is
 \beq\label{eq:C3}
(2i\mu\omega_n-\mu^2)\delta Z+\frac{\delta\lambda}{3} \Big[{\cal T}[{\cal D}]-{\cal T}[{\cal D}_{\mu=0}]\Big].
\eeq
Since we deal with differences of diagrams involving either the propagators ${\cal D}$ and ${\cal D}^*$, or the propagator ${\cal D}_{\mu=0}={\cal D}_{\mu=0}^*$, it is now natural to write ${\cal D}$ as ${\cal D}_{\mu=0}+\Delta {\cal D}$ and ${\cal D}^*$ as ${\cal D}_{\mu=0}+\Delta {\cal D}^*$. Plugging these decompositions in the sunset difference, one obtains $8$ sunset type integrals, with one, two or three occurrences of $\Delta{\cal D}$ (which can also appear in the form $\Delta {\cal D}^*$). In order to discuss the UV behaviour of each of these pieces, it is convenient to extract the leading UV contributions to $\Delta {\cal D}$:
\beq\label{eq:dec}
\Delta {\cal D} & = & {\cal D}-{\cal D}_{\mu=0}\nonumber\\
 & = & (\mu^2-2\mu i\omega_n){\cal D}{\cal D}_{\mu=0}\nonumber\\
& = & (\mu^2-2\mu i\omega_n){\cal D}^2_{\mu=0}+(\mu^2-2\mu i\omega_n)^2{\cal D}^2{\cal D}_{\mu=0}\nonumber\\
& = & (\mu^2-2\mu i\omega_n){\cal D}^2_{\mu=0}-4\mu^2 \omega_n^2{\cal D}^3_{\mu=0}+{\cal D}_r\,.
\eeq
From this expansion, it is clear for instance that the sunset contribution with three occurrences of $\Delta {\cal D}$ is convergent by simple power counting. For those contributions involving two occurrences of $\Delta {\cal D}$, there are clearly no subdivergences, because any subloop involves at least one $\Delta {\cal D}$ which decreases the superficial degree of divergence of the original logarithmically divergent subloop. Thus, in the contributions involving two occurrences of $\Delta {\cal D}$ there can only be an overall divergence which we shall discuss more precisely in a moment. Finally, those contributions involving one occurrence of $\Delta {\cal D}$ rewrite $(-\lambda^2/18)({\cal S}[\Delta{\cal D},{\cal D}_{\mu=0},{\cal D}_{\mu=0}]+{\cal S}[{\cal D}_{\mu=0},\Delta{\cal D}^*,{\cal D}_{\mu=0}]+{\cal S}[{\cal D}_{\mu=0},{\cal D}_{\mu=0},\Delta{\cal D}])$. They involve both subdivergences and overall divergences. However, when combined with the second counterterm diagram in (\ref{eq:C3}), which we rewrite for convenience as $(\delta\lambda/9) ({\cal T}[\Delta{\cal D}]+{\cal T}[\Delta{\cal D}^*]+{\cal T}[\Delta{\cal D}])$, it is pretty clear that what remains are again overall divergences. These, together with those present in the terms with two occurrences of $\Delta {\cal D}$ can be treated in the following way. We note first that, owing to (\ref{eq:dec}), it is only logarithmic and thus originates from the zero temperature contribution of the diagram. Then, from the Silver Blaze property, we conclude that the overall divergence has the same structure than the one at $\mu=0$, but with $\omega_n$ replaced by $\omega_n+i\mu$ ($+i\mu$ here because we are considering a contribution to the correlator $\langle\Phi^*(x)\rangle\Phi(y)$), namely it is proportional to $K^2=(\omega_n+i\mu)^2+k^2$. But this is precisely the same structure than the first counterterm in (\ref{eq:C3}).\\

The discussion in the general case follows a similar argumentation. The recursive way of renormalizing a
diagram finds always the smallest overall divergent subgraph (or a subgraph
together with its subdivergences already accounted for) of a graph and adds
a diagram where the divergent subgraph is replaced by a counterterm. Because
of the Silver Blaze property coupling and mass overall divergences are unchanged as compared to
their $\mu=0$ value, whereas field-strength divergences are modified just because they are proportional to $(\omega_n+i\mu)^2+k^2$ (or $(\omega_n-i\mu)^2+k^2$ depending on the type of self-energy insertion one is looking at) rather than $\omega^2_n+k^2$ but the divergent factor that multiplies these quadratic functions remains independent of $\mu$.

\subsection{Renormalization of the density}
Because renormalization of $\Phi$-derivable approximations is just a resummed version of perturbative renormalization, the above discussion is enough to argue that $\Phi$-derivable approximations at finite $\mu$ are renormalized by exactly the same counterterms than those needed at $\mu=0$. In fact, extending the considerations of the above section it is in fact possible to show explicitly that the bare parameters of \cite{Marko:2013lxa} together with the renormalization factor $Z_2$ given in (\ref{eq:Z2Explicit}) are enough to renormalize the two-loop $\Phi$-derivable approximation at finite $\mu$. We shall not do this here in full glory. Instead we concentrate on the renormalization of the density in the broken phase which requires the use of $Z_2$.

The density in the broken phase is given by
\beq\label{eq:rhoSep}
\rho=\mu Z_2\bar\phi^2+{\cal T}[\mu(\bar G_L+\bar G_T)-2\omega_n \bar G_A]\,.
\eeq
We now use a UV expansion of the propagators:
 \beq\label{eq:lead1}
\bar G_{L,T}=G_\star-\Delta\bar M^2_{L,T}G^2_\star-4\mu^2\omega^2_n G^3_\star+\dots
\eeq
 and
 \beq\label{eq:dots}
\omega_n\bar G_A & = & -\omega_n\bar M^2_AG_\star^2-2\mu\omega^2_nG^3_\star(\Delta\bar M^2_L+\Delta \bar M^2_T)\nonumber\\
& & +\,2\mu\omega^2_nG_\star\Big(G_\star-4\mu^2\omega^2_nG^3_\star\Big)+\dots
\eeq
where $G_\star=1/(Q^2+m^2_\star)$ is a reference massive free propagator, $\Delta \bar M_{L,T}^2=\bar M_{L,T}^2-m_\star^2-\mu^2$ and the dots denote terms which do not lead to divergences in $\rho$. Moreover, we need only the dominant large momentum behavior of $\Delta\bar M^2_{L,T}$ and $\bar M^2_A$ which up to subleading terms reads
\beq
\Delta\bar M^2_L & = & \Delta\bar M^2_{L,l}-\frac{5\lambda_\star}{36}\phi^2\left[{\cal B}[G_\star](K)-{\cal B}[G_\star](0)\right]+\dots,\label{eq:lead3}\nonumber\\\\
\Delta\bar M^2_T & = & \Delta\bar M^2_{T,l}-\frac{\lambda_\star}{36}\phi^2\left[{\cal B}[G_\star](K)-{\cal B}[G_\star](0)\right]+\dots,\label{eq:lead4}\nonumber\\
\eeq
and
\beq
\bar M_A^2= \mu\phi^2\frac{\lambda_\star^2}{9}{\cal B}[\omega G_\star^2;G_\star](K)+\dots\label{eq:lead5}
\label{eq:MA2LO}
\eeq
where $\Delta M^2_{L,T,l}$ denote momentum independent (local) UV finite contributions which we do not need to specify more. Plugging (\ref{eq:lead1}) and (\ref{eq:dots}), with (\ref{eq:lead3}), (\ref{eq:lead4}) and (\ref{eq:lead5}), back into \eqref{eq:rhoSep}, the potentially divergent contributions to $\rho$ are
\beq\label{eq:terms}
&&\mu \phi^2\left(\frac{2\lambda^2_\star}{9}\int_Q^T\int^T_K\omega_nG_\star^2(Q)\omega_m G^2_\star(K)G_\star(K+Q)\right.\nonumber\\
&&\hspace{1.0cm}\left.+\,\frac{\lambda_\star^2}{6}\int_Q^T \big(G^2_\star-4\omega^2_nG^3_\star\big)\big[{\cal B}[G_\star](Q)-{\cal B}[G_\star](0)\big]\right)\nonumber\\
&&+\mu\left(\,2\int_Q^T G_\star(1-2\omega_n^2G_\star)-8\mu^2\int_Q^T\omega_n^2G_\star^3(1-2\omega_n^2G_\star) \right.\nonumber\\
&&\hspace{1.0cm}\left.-\,(\Delta\bar M_{L;l}^2+\Delta\bar M_{T;l}^2)\int_Q^TG_\star^2(1-4\omega_n^2G_\star)\right).
\label{eq:rho1stLineFin}
\eeq
The last three integrals present in \eqref{eq:rho1stLineFin} are finite, as one checks by direct calculations of the Matsubara sums. To ease these calculations, it is convenient to express all the Matsubara sums in terms of the tadpole one. For instance, we write
\beq
\int_Q G_\star(1-2\omega^2_nG_\star) & = & \int_Q\left(-G_\star+2\varepsilon_q^2G_\star^2\right)\nonumber\\
& = & \int_Q\left(-1-2\varepsilon_q^2\frac{d}{dq^2}\right)G_\star\nonumber\\
& = & \int_q\underbrace{\left(-1-2\varepsilon^2_q\frac{d}{dq^2}\right)\frac{1}{2\varepsilon_q}}_{=0}+{\rm finite}\,.\nonumber\\
\eeq
Similarly
\beq
& & \int_Q^T G^2_\star(1-4\omega^2_nG_\star)\nonumber\\
& & \hspace{0.5cm}=\,\int_Q^T (-3G^2_\star+4\varepsilon^2_q G^3_\star)\nonumber\\
& & \hspace{0.5cm}=\,\int_Q^T\left(3\frac{d}{dq^2}+2\varepsilon_q^2\frac{d^2}{d(q^2)^2}\right)G_\star\nonumber\\
& & \hspace{0.5cm}=\,\int_q\underbrace{\left(3\frac{d}{dq^2}+2\varepsilon_q^2\frac{d^2}{d(q^2)^2}\right)\frac{1}{2\varepsilon_q}}_{=0}+{\rm finite}\,.\nonumber\\
\eeq
Finally
\beq
& &\int_Q^T\omega^2_n G^3_\star(1-2\omega^2_nG_\star)\nonumber\\
& & \hspace{0.5cm}=\,\int_Q^T\omega^2_n\left(-G^3_\star+2\varepsilon_q^2G^4_\star\right)\nonumber\\
& & \hspace{0.5cm}=\,\int_Q^T\left(-G^2_\star+3\varepsilon_q^2G^3_\star-2\varepsilon_q^4G^4_\star\right)\nonumber\\
& & \hspace{0.5cm}=\,\int_Q^T\left(\frac{d}{dq^2}+\frac{3}{2}\varepsilon_q^2\frac{d^2}{d(q^2)^2}+\frac{1}{3}\varepsilon_q^4\frac{d^3}{d(q^2)^3}\right)G_\star\nonumber\\
& & \hspace{0.5cm}=\,\int_q\underbrace{\left(\frac{d}{dq^2}+\frac{3}{2}\varepsilon_q^2\frac{d^2}{d(q^2)^2}+\frac{1}{3}\varepsilon_q^4\frac{d^3}{d(q^2)^3}\right)\frac{1}{2\varepsilon_q}}_{=0}\nonumber\\
& & \hspace{0.5cm}+\,{\rm finite}\,.
\eeq
In order to discuss the first two lines of (\ref{eq:terms}), we note first that they only contain an overall and thus temperature-independent divergence. Then, a comparison to \eqref{eq:Z2Explicit} shows that this overall divergence is exactly cancelled by $Z_2$. This completes the proof of the finiteness of the density.

\section{Sum-integrals}\label{app:sum}
In this section we compute certain sum-integrals which occur in the symmetric phase, in particular in the equation defining the transition line. For simplicity we consider dimensional regularisation but the calculation can be easily adapted to any regularization that does not cut off the Matsubara sums. Let us consider first the tadpole sum-integral
\beq\label{eq:ff}
{\cal T}[{\cal D}]=\int_Q^T {\cal D}(Q)
\eeq
with ${\cal D}(Q)=1/((\omega_n+i\mu)^2+q^2+m^2)$ and $|\mu|<m$. To compute the Matsubara sum, we use the formula
\beq\label{eq:mat}
T\sum_{n=-\infty}^{+\infty}f(i\omega_n)=-\sum_{\mbox{\tiny p\^oles $z$ of f}} n(z)\,{\rm Res}\,f(z)
\eeq
valid for any function $f$ whose p\^oles are simple (and not equal to any $i\omega_n$). Because
\beq\label{eq:dd}
{\cal D}(Q)=-\frac{1}{2\varepsilon_q}\left[\frac{1}{i\omega_n-\mu-\varepsilon_q}-\frac{1}{i\omega_n-\mu+\varepsilon_q}\right],
\eeq
it follows from (\ref{eq:mat}) that
\beq
{\cal T}[{\cal D}] & = & \int_q\frac{1}{2\varepsilon_q}\Big[n_{\varepsilon_q+\mu}-n_{-\varepsilon_q+\mu}\Big]\nonumber\\
& = & \int_q\frac{1}{2\varepsilon_q}\Big[1+n_{\varepsilon_q-\mu}+n_{\varepsilon_q+\mu}\Big],
\eeq
where we have used $n_x=-1-n_{-x}$. We note that the zero temperature contribution to ${\cal T}[{\cal D}]$ is $\mu$-independent. We see here at play the Silver Blazer property or more precisely its generalization to vertex functions. At zero temperature and in the symmetric phase, the $\mu$-dependence of vertex functions is trivial and amounts to appropriate shifts of the external frequencies of the corresponding vertices at $\mu=0$. Since the tadpole does not depend on the external frequency, its zero temperature contribution cannot depend on $\mu$. This can also be checked without performing any Matsubara sum. The zero temperature limit of (\ref{eq:ff}) writes
\beq
{\cal T}_{T=0}[{\cal D}]=\int_Q^{T=0}\frac{1}{(\omega+i\mu)^2+q^2+m^2}
\eeq
where the Matsubara sum has been replaced by an integral $\int_{\mathds{R}}d\omega$. But because $|\mu|<m$, we can deform the contour of integration to $-i\mu+\mathds{R}$, without encountering any singularity and thus
\beq
{\cal T}_{T=0}[{\cal D}]=\int_Q^{T=0}\frac{1}{\omega^2+q^2+m^2}\,,
\eeq
which shows that ${\cal T}_{T=0}[{\cal D}]$ is independent of $\mu$.\\

We next compute the sunset sum-integral
\beq\label{eq:sss}
{\cal S}[{\cal D};{\cal D}^*;{\cal D}]=\int_Q^T\int_K^T {\cal D}(Q){\cal D}^*(K){\cal D}^*(-K-Q)\,.\nonumber\\
\eeq
To this purpose, we follow the lines of \cite{Blaizot:2004bg}. It is convenient to use the spectral representations\footnote{It could be tempting to remove $\mu$ from the denominator appearing in the spectral representation by including it in the spectral function. The calculations are however more straightforward if one leaves $\mu$ in the denominator.}
\beq
{\cal D}(Q) & = & \int_Q\frac{\rho(q_0,q)}{q_0-i\omega_n+\mu}\,,\\
{\cal D}^*(Q) & = & \int_Q\frac{\rho(q_0,q)}{q_0+i\omega_n+\mu}=\int_Q\frac{\rho(q_0,q)}{q_0-i\omega_n-\mu}\,,
\eeq
which follow from (\ref{eq:dd}), with
\beq
\rho(q_0,q) & = & \frac{1}{2\varepsilon_q}\Big[\delta(q_0-\varepsilon_q)-\delta(q_0+\varepsilon_q)\Big]=-\rho(-q_0,q)\,.\nonumber\\
\eeq
Performing the double Matsubara sum and using the identity $(1+n_x+n_y)n_{x+y}=n_x n_y$, we arrive at
\begin{widetext}
\beq\label{eq:aa}
{\cal S}[{\cal D};{\cal D}^*;{\cal D}] & = & \int_{q_0,q}\int_{k_0,k}\int_{p_0,p}(2\pi)^{d-1}\delta^{(d-1)}(\vec{q}+\vec{k}+\vec{p})\rho(q_0,q)\rho(k_0,k)\rho(p_0,p)\nonumber\\
& & \hspace{2.5cm}\times\,\frac{n_{q_0+\mu}n_{p_0-\mu}+n_{q_0+\mu}(-n_{-k_0+\mu})+(-n_{-p_0+\mu})(-n_{-k_0+\mu})}{q_0+k_0+p_0-\mu+i\alpha}\,.
\eeq
Because $q_0=\pm\varepsilon_q$ due to the spectral function, $|q_0|>\mu$ and then the identities $n_{q_0+\mu}=-\theta(-q_0)+\varepsilon(q_0)n_{|q_0|+\varepsilon(q_0)\mu}$ and $-n_{-q_0+\mu}=\theta(q_0)+\varepsilon(q_0)n_{|q_0|-\varepsilon(q_0)\mu}$ split the thermal factors into a zero temperature contribution and a contribution which vanishes as $T\to 0$. Plugging these decompositions in (\ref{eq:aa}), we arrive at
\beq\label{eq:formule}
{\cal S}[{\cal D};{\cal D}^*;{\cal D}] & = & {\cal S}_{T=0}[{\cal D}_{\mu=0}](P_\mu)\nonumber\\
 & + & \int_{q_0}\int_q \varepsilon(q_0)\rho(q_0,q)(n_{|q_0|+\varepsilon(q_0)\mu}+2n_{|q_0|-\varepsilon(q_0)\mu}){\cal B}_{T=0}[{\cal D}_{\mu=0}](q_0-\mu+i\alpha,q)\nonumber\\
& + & \int_{q_0}\int_q \int_{k_0}\int_k \Big[\varepsilon(q_0)\rho(q_0,q)\varepsilon(k_0)\sigma(k_0,k)(n_{|q_0|-\varepsilon(q_0)\mu}+2n_{|q_0|+\varepsilon(q_0)\mu})n_{|k_0|-\varepsilon(k_0)\mu}\nonumber\\
&&\hspace{2.1cm}\times{\cal D}_{\mu=0}(q_0+k_0-\mu+i\alpha,|\vec{q}+\vec{k}|)\Big]\,,
\eeq
\end{widetext}
where we have made use of the spectral representation and we have identified the analytic continuation ${\cal B}_{T=0}[{\cal D}_{\mu=0}](q_0-\mu+i\alpha,q)$ of the zero temperature integral
\beq
{\cal B}_{T=0}[{\cal D}_{\mu=0}](K)=\int_Q^{T=0} {\cal D}_{\mu=0}(Q){\cal D}_{\mu=0}(Q+K)\,.\nonumber\\
\eeq
Moreover ${\cal S}_{T=0}[{\cal D}_{\mu=0}](P_\mu)$ stands for the zero temperature sunset integral
\beq\label{eq:above}
\int_Q^{T=0}\int_K^{T=0}{\cal D}_{\mu=0}(Q){\cal D}_{\mu=0}(K){\cal D}_{\mu=0}(K+Q+P_\mu)\nonumber\\
\eeq
where $P_\mu=(i\mu,0)$. Because the thermal distribution functions in the last two lines of (\ref{eq:formule}) vanish in the limit $T\to 0$, the above expression (\ref{eq:above}) is nothing but the value of the setting-sun diagram at finite chemical potential ($|\mu|<m$) and zero temperature. This result is once again an illustration of the Silver Blaze property. The zero temperature contribution to the sunset, which from (\ref{eq:sss}) writes
\beq
& & \int_Q^{T=0}\int_K^{T=0}\frac{1}{(\omega+i\mu)^2+\varepsilon_q^2}\frac{1}{(\nu-i\mu)^2+\varepsilon_k^2}\nonumber\\
& & \hspace{3.0cm}\times\,\frac{1}{(\nu+\omega+i\mu)^2+\varepsilon_{\vec{k}+\vec{q}}^2}
\eeq
is in fact equal to the corresponding integral at $\mu=0$ but with the external frequency shifted by $i\mu$:
\beq
\int_Q^{T=0}\int_K^{T=0}\frac{1}{\omega^2+\varepsilon_q^2}\frac{1}{\nu^2+\varepsilon_k^2}\frac{1}{(\nu+\omega+i\mu)^2+\varepsilon_{\vec{k}+\vec{q}}^2}\nonumber\\\nonumber\\
\eeq
as it can be checked more directly by noticing that one can deform the contours of integration from $\omega\in\mathds{R}$ and $\nu\mathds{R}$ to $\omega\in -i\mu+\mathds{R}$ and $\nu\in i\mu+\mathds{R}$ without encountering any singularity.

Finally, integrating over the frequencies and performing trivial angular integrals, we arrive at
\begin{widetext}
\beq
{\cal S}[{\cal D};{\cal D}^*;{\cal D}] & = & {\cal S}_{T=0}[{\cal D}_{\mu=0}](P_\mu)\nonumber\\
 & + & \frac{1}{4\pi^2}\int_0^\infty dq\,\frac{q^2}{\varepsilon_q}\sum_\sigma (n_{\varepsilon_q+\sigma\mu}+2n_{\varepsilon_q-\sigma\mu})\,{\cal B}_{T=0}[{\cal D}_{\mu=0}](\sigma\varepsilon_q-\mu+i\alpha,q)\Big]\nonumber\\
& + & \frac{1}{16\pi^3}\int_0^\infty dq \frac{q}{\varepsilon_q} \int_0^\infty dk\,\frac{k}{\varepsilon_k}\sum_{\sigma,\tau} (n_{\varepsilon_q-\sigma\mu}+2n_{\varepsilon_q+\sigma\mu})n_{\varepsilon_k-\tau\mu}\nonumber\\
 & & \hspace{3.0cm}\ln\frac{-(\sigma\varepsilon_q+\tau\varepsilon_k-\mu+i\alpha)^2+(q+k)^2+\bar M^2}{-(\sigma\varepsilon_q+\tau\varepsilon_k-\mu+i\alpha)^2+(q-k)^2+\bar M^2}\,.
\eeq
\end{widetext}
In dimensional regularization, the zero temperature integrals ${\cal S}_{T=0}[{\cal D}_{\mu=0}](P_\mu)$ and ${\cal B}_{T=0}[{\cal D}_{\mu=0}](\sigma\varepsilon_q-\mu+i\alpha,q)$ can be evaluated using standard techniques.


\end{document}